\documentclass[%
 aip,
 jmp,%
 amsmath,amssymb,
 reprint,%
author-year,%
]{revtex4-1}
\pdfoutput=1

\usepackage{graphicx}
\usepackage{dcolumn}
\usepackage{bm}

\def\vec#1{{\mbox{\boldmath{$#1$}}}}

\def\vp{\vec{\times}}
\def\sp{\vec{\cdot}}

\def\cl{:\ }
\def\scl{;\ }
\def\vlne{|}

\def\del#1#2{\frac{\partial \mbox{$#1$}}{\partial {\mbox{$#2$}}}}

\def\dif#1#2{ \frac{d\mbox{$#1$}}{d\mbox{$#2$}}}

\def\vnab{\vec{\nabla}}

\def\lini{\lambda_{\rm i}}

\def\ocyci{\Omega_{{\rm ci}}}

\def\nion{n_{\rm{i}}}
\def\nel{n_{\rm{e}}}

\def\me{m_{\rm e}}
\def\mi{m_{\rm i}}

\def\ic{\frac{1}{c}}

\def\apj{{\rm Astrophys. J}}
\def\nat{{\rm Nature}}

\begin{document}

\preprint{AIP/123-QED}

\title[Magnetic Reconnection in Collisionless Accretion disk.]{Asymmetric Evolution of Magnetic
Reconnection in Collisionless Accretion Disk}

\author{Keisuke Shirakawa}
 \email{k-suke@eps.s.u-tokyo.ac.jp}
\author{Masahiro Hoshino}%
\affiliation{Department of Earth and Planetary Science, The University of Tokyo, 7-3-1 Hongo, Bunkyo-ku, Tokyo, Japan}%



\date{\today}



\begin{abstract}
 An evolution of a magnetic reconnection in a  
 collisionless accretion disk is investigated using a 2.5 dimensional
 hybrid code simulation. In astrophysical disks magnetorotational instability
 (MRI) is considered to play an important role by generating turbulence
 in the disk and contributes to an effective angular momentum transport
 through a turbulent viscosity. Magnetic reconnection, on the other hand
 also plays an important role on the evolution of the disk through a 
 dissipation of a magnetic
 field enhanced by a dynamo effect of MRI. In this study, we developed
 a hybrid code to calculate an evolution of a differentially rotating
 system.
 With this code, we first confirmed a linear growth of MRI. We also
 investigated a behavior of a particular structure of a current sheet
 which would exist in the turbulence in the disk. From the calculation
 of the magnetic reconnection, we found an asymmetric structure in the
 out-of-plane magnetic field during the evolution of reconnection which
 can be  understood by a coupling of the Hall effect and the differential
 rotation. We also found a migration of X-point whose direction is
 determined only by an initial sign of $\vec{J}_{0} \vec{\times
 \Omega}_{\rm 0}$ where $\vec{J}_{\rm 0}$ is the initial current density
 in the neutral sheet and $\vec{ \Omega}_{0}$ is the rotational vector
 of the background Keplerian rotation.
 Associated with the migration of X-point we also found a significant
 enhancement of the perpendicular magnetic field compared to an ordinary
 MRI. MRI-Magnetic reconnection coupling and the resulting
 magnetic field enhancement can be an effective process to sustain a
 strong turbulence in the accretion disk and to a transport of angular
 momentum.
 

%
\end{abstract}


\keywords{Magnetic Reconnection, Accretion Disk, Magneto Rotational
Instability, Collisionless Plasma}
\maketitle



\section{Introduction}\label{sec_introduction}
Transport of angular momentum in accretion disks has been one of the
fundamental problems in astrophysics. For the efficient transport of
the angular momentum, turbulence generated in a disk is believed to
play an important role through strong turbulent viscosity
[\cite{ShakuraSunyaev1973}]. The most promising process to generate
turbulence in accretion disks is the magnetorotational instability (MRI) 
which is driven by a strong dynamo effect of a shear flow provided by
the differential rotation of the disk
[\cite{Velikhov1959}\scl\cite{Chandrasekhar1960}\scl\cite{BalbusHawley1991A}]. Since
the turbulence in the disk determines the efficiency of the angular
momentum transport, saturation of the MRI has been one of the most
important problems. In the saturated, quasi steady state, the dynamo effect
of the MRI and some sort of magnetic dissipation must be
balanced [\cite{BH1998Review}]. 
Numerous simulation studies have been proposed by many researchers
to study the behavior of a quasi steady state of the MRI induced
turbulence [\cite{BalbusHawley1991B}\scl\cite{BalbusHawley1991C}\scl\cite{MatsumotoTajima1995}\scl\cite{SanoStone2002A}].
During the turbulent state of those calculations rapid amplification
and dissipation of magnetic energy have been observed implying
magnetic reconnection is playing an important role on the dissipation of
magnetic energy.

Until now, behavior of the plasma in accretion disks is mainly studied
based on a conventional MHD approximation. However, in some classes of
accretion
disks a mean free path of the plasma is estimated to be comparable to
the scale of the disk thickness. For example, several observations
suggest that in the accretion disk around
Sgr A$^{\ast}$ at the center of our galaxy the ions and electrons are
not in the thermal equilibrium. Since the ions are much hotter than the
electrons, the plasma constituting the disk is considered to be
collisionless. In such a disk kinetic
effect of the plasma is considered to be important since a pressure
anisotropy generated by the stretching of the magnetic field modifies
a feature of magnetic tension. 

Based on these facts, several attempts to understand both linear and
nonlinear evolution of collisionless MRI have been
made. \cite{Quataert+2002} have studied a linear behavior
of MRI under CGL approximation
[\cite{CGL1956}\scl\cite{Kulsrud1983Text}] together with parallel heat
flux due to the Landau damping incorporated via the so-called Landau
closure [\cite{Snyder1997}]. They have shown that the
pressure anisotropy modifies the linear behavior of the MRI and have
suggested 
that the kinetic plasma effect would also be important in the nonlinear
evolution
of MRI. \cite{Sharma+2006} have extended this study to a nonlinear
regime by using a so-called collisionless MHD code. Based on the
linear analysis, they
have applied the CGL approximation together with the Landau closure for
calculating the temporal evolution of the parallel and the perpendicular
pressure with respect to the local magnetic field. In addition, they
have assumed that the
upper and the lower limit of the pressure anisotropy is limited by the
criteria of the firehose, the mirror and the ion cyclotron
instabilities. These
instabilities would cause pitch angle scattering of the ions and lead to a
relaxation of the generated pressure anisotropy. They have found that
with the
equation of state based only on the standard CGL approximation,
enhancement of the magnetic field generates the perpendicular pressure
anisotropy and have obtained a remarkably low saturation level of the MRI.
The result was well understood as a suppression of the MRI by the
generated perpendicular pressure anisotropy, because the perpendicular
pressure anisotropy effectively enhances the magnetic tension which act
as the restoring force in the evolution of the MRI.
 With the parallel heat flux and the appropriate pitch angle
scattering model, they confirmed the nonlinear evolution of the MRI and
found that the rate of the angular momentum transport is enhanced
moderately.
 
Numerical simulations of MRI capturing full kinetic effects in a
consistent manner have been performed by \cite{Riquelme+2012} and
\cite{Hoshino2013} by using a 2.5D PIC code which treats both the ions
(the positrons, in \cite{Hoshino2013}) and the electrons as superparticles.
\cite{Riquelme+2012} have shown that during the evolution of the
kinetic MRI, the 
pressure anisotropy has been indeed, relaxed by the mirror mode and
have confirmed basic features in the nonlinear stage of the MRI
reported in \cite{Sharma+2006}. They have also found that the channel
flow and the current
sheet which are usually found in the final state of the 2.5D MHD
simulations were corrupted due to a
magnetic reconnection. \cite{Hoshino2013} has extended the study to a
relatively high beta ($\beta\simeq$90-6000) pair plasma and
found the corruption of channel flow as well. Moreover they found
that during the ``active'' phase when the magnetic reconnection takes
place repeatedly and the channel flow is corrupted, the Maxwell stress is
greatly enhanced. In both simulations particle acceleration associated
with the magnetic reconnection has also been observed.

Kinetic effect of plasma is also considered to be important in the
evolution of magnetic reconnection. 
Several {\it in-situ} observations of the earth's magnetotail,
which is also constituted by a collisionless plasma, have
pointed out the existence of thin current sheet [\cite{Asano+2003}].
In those thin current sheets, the thickness has found to be 
comparable to the scale of the ion inertia length.
In the magnetic reconnection with such a thin current sheet
effect of the Hall term in the generalized Ohm's law is also important
[\cite{Sonnerup1979Text}\scl\cite{Terasawa1983}]. Therefore, for the
understanding of the magnetic reconnection in the
collisionless accretion disks, the kinetic effect of the plasma, together
with the effect of the differential rotation should be taken into
account in a self-consistent manner.

For an investigation of kinetic effect of the plasma, PIC
simulation gives a self-consistent picture. However
the attainable scale of the method is roughly limited in a
several hundred times of the Debye length which is far smaller than that
 of the actual disk. In addition, as full particle code must resolve
a time scale related to the electron physics it requires a massive
integration time if one tries to focus both on the Keplerian time scale
and on the
electron time scale. Moreover, under a ``first order'' approximation
kinetic effect of the ions
should be more important than that of the electrons since it is often found
in a collisionless accretion disk, the thermal energy of the ions
dominates that of the electrons. In this point of view, a hybrid code which
treats the ions as particles and the electrons as massless fluid would
provide a more robust approach rather than a full particle method. 

In this work, we will study an evolution of a differentially rotating
system with a hybrid code including the effect of the Coriolis and the
tidal force. 
As a first step, we will mainly focus on an evolution of a magnetic
reconnection in the differentially rotating system, which would exist in
the turbulence generated by the MRI. The structure of this article is as
follows. In
Sec. \ref{sec_simulation}, we first describe our
2.5D hybrid code which includes the Coriolis and the tidal force with an
open shearing boundary condition [\cite{Hawley+1995}]. Since
our simulation here is focused on a 2.5D system, the boundary
condition degenerates to a conventional periodic boundary condition for 
the deviation components from the background differential rotation. Next in
Sec. \ref{sec_results}, we introduce the results of the five
consecutive runs. We also briefly introduce the linear growth of the MRI
calculated with our hybrid code. At last in Sec. \ref{sec_dis}, we
summarize our results and discuss their effects on the MRI.






\section{Hybrid Simulation in a Differentially Rotating
 System}\label{sec_simulation}

We focus on a time evolution of the magnetic reconnection in a localized
region of an accretion disk. Under the so-called 'local approximation', a
number of simulations have been carried out to study a nonlinear
evolution of the
system under the differential rotation. Those studies which mainly
focused on the evolution of the MRI are performed basically under the MHD
approximation
[\cite{BalbusHawley1991B}\scl\cite{BalbusHawley1991C}\scl\cite{MatsumotoTajima1995}\scl\cite{SanoStone2002A}],
and recently using PIC codes [\cite{Riquelme+2012}\scl\cite{Hoshino2013}],
taking kinetic plasma effect into 
account in a self-consistent manner. Basic equations of our simulation
are essentially the same with those used in the previous studies, but by
using a hybrid code, we took kinetic effect of the ions into account in
a self-consistent manner and covered spatially wider region compared to
those calculated in PIC simulations. Like in the MHD and the PIC
calculations, one must pay
attention to the boundary conditions in the differentially rotating
system. A brief introduction about the boundary condition is given
in Appendix \ref{appendix_sh_bnd}.

\subsection{Basic Equations}\label{subsec_sim_method}
As we focus on a local evolution of the magnetic reconnection, we
describe a set of equations in a local corotating frame with the angular
velocity of $\vec{\Omega}_{0}$ at the distance of $r_{0}$ from the central
massive object.
In this rotating coordinate, we solve a set of Lorentz-Maxwell
equations. The ions are treated as superparticles and their motion is
calculated with the Lorentz equation including the Coriolis force,
the centrifugal force, and gravity, expressed with the cylindrical
representation ($r$,$\varphi$,$z$),
\begin{eqnarray}
 \mi \dif{\vec{v}_{\rm i}}{t} &=& e \left( \vec{E} + \ic \vec{v}_{\rm i}
				   \vp \vec{B} \right) 
 - 2 \mi \vec{\Omega}_{0} \vp \vec{v}_{\rm i} + \mi r_{\rm i}
 \Omega_{0}^{2} \vec{e}_{r} - \mi g(\vec{r}_{\rm i}) \vec{e}_{r}
 ,\label{eq_eom_cyl_v}
 \\
  \dif{\vec{r}_{\rm i}}{t} &=& \vec{v}_{\rm i}.   \label{eq_eom_cyl_x}
\end{eqnarray}
Here $\mi$ is the ion
mass, $\vec{v}_{\rm i}$ is the ion velocity, $\vec{r}_{\rm i}$ is the ion
position, $e$ is the ion charge, $c$ is the speed of light, $\vec{E}$ is the
electric field,  $\vec{B}$ is the magnetic field, $g(\vec{r}_{\rm
i})$ is the gravity from the central object, and $\vec{e}_{r}$ is an unit
vector of $r$-component. Throughout this paper the vertical component
($z$-direction) of the gravity is ignored, and in the
background rotation, balance between the gravity and the centrifugal force,
$\mi g(\vec{r}_{\rm i}) = \mi r \Omega_{0}^{2}$ is satisfied. 

The electron, on the other hand is assumed to be a massless charge
neutralizing fluid, and a generalized Ohm's law is given by
\begin{eqnarray}
 \vec{E} = - \ic \vec{v}_{\rm e} \vp \vec{B} + \frac{1}{\nel e} \vnab{p_{\rm
  e}} + \eta \vec{J}, \label{eq_Ohm}
\end{eqnarray}
where $\nel=\nion$ is the number density of the electrons (ions),
$\vec{v}_{\rm e}$
is the moment averaged velocity of the electron, $\eta$ is the
resistivity and $\vec{J}$ is the current density.
Since we focus on the ion physics and assume the kinetic processes of
the electrons,
such as pitch angle scattering would take place much faster than those
of the ions, we take the electron pressure, $p_{\rm e}$ as a scalar variable and
assume to be given by an adiabatic relation,
\begin{eqnarray}
 p_{\rm e} \propto n_{\rm e}^{\gamma}, \label{eq_Eadia}
\end{eqnarray}
where $\gamma=5/3$ is the adiabatic constant.
In addition, the Coriolis force, the centrifugal force and the gravity are
neglected for the electrons because of a limit $\me \rightarrow 0$.
Finally, with the Maxwell equations under the Darwin approximation,
\begin{eqnarray}
 \vnab \vec{\cdot B} &=& 0, \label{eq_solenoidal} \\
 \vec{J} &=& \vec{J}_{\rm e} + \vec{J}_{\rm i} = \frac{c}{4 \pi} \vnab
  \vp \vec{B},   \label{eq_Ampere} \\ 
 \del{\vec{B}}{t} &=& -c \vnab \vp \vec{E}, \label{eq_induction}
\end{eqnarray}
we obtain a closed set of equations.

In this study we investigate the evolution of the system in the
meridional plane $(r,z)$, of the accretion disk.
Since we focus on the localized region of the accretion disk, we 
approximate the background differential rotation with a linear
profile. Introducing the following transformation,
\begin{eqnarray}
 x &=& r - r_{0}, \label{eq_Hillx}\\
 z &=& z, \label{eq_Hillz}
\end{eqnarray}
and applying the tidal expansion on the Hill coordinate [\cite{Hill1878}],
 the equation of motion (\ref{eq_eom_cyl_v}) becomes,
\begin{eqnarray}
 \mi \dif{\vec{v}_{\rm i}}{t} = e \left( \vec{E} + \ic \vec{v}_{\rm i}
				   \vp \vec{B} \right) 
 - 2 \mi \vec{\Omega}_{0} \vp \vec{v}_{\rm i}
 -2 \mi q \Omega^{2}_{0} x \vec{e}_{x},  \label{eq_tidal_eom}
\end{eqnarray}
where $\vec{e}_{x}$ is an unit vector of $x-$component and  $q =
\partial \ln \Omega / \partial \ln r |_{r = r_{0}}$ and for
a Keplerian rotation $ \left( g(\vec{r}_{\rm i}) \propto r^{-2} \right)$,
$q$ is equal to $-3/2$. Note that as we assume the size of the
simulation domain is much smaller compared to $r_{0}$, the above
equations may now be expressed in the Cartesian coordinate.
Under this approximation, we can write the background Keplerian velocity in
the corotating frame as,
\begin{eqnarray}
 \vec{v}_{\rm K} = q \Omega_{0} x \vec{e}_{y}, \label{eq_vkepler}
\end{eqnarray}
where $\vec{e}_{y}$ is an unit vector of $y$-component.
Because of this background flow, non-zero electric field $\vec{E}_{\rm
K} = - \vec{v}_{\rm K} \vp \vec{B}/c$ is always observed in the
corotating frame. Since this electric field is a function of $x$,
charge neutrality is not exactly satisfied in this system. The magnitude
of the charge inequality is estimated as,
\begin{eqnarray}
 \frac{ \vnab \sp \vec{E}_{\rm K} }{n_{\rm i0} e } &=& - \frac{1}{n_{\rm
  i0} e c } 
  q \Omega_{0} B  - \frac{1}{n_{\rm i0} e c }
  q \Omega_{0} B \left. \del{\ln B}{\ln x} \right|_{x = 0}  \nonumber \\
  &=& - 4 \pi \frac{V_{\rm A}^{2}}{c^{2}} \frac{\Omega_{0}}{\ocyci}
   \left( 1 + \left. \del{\ln B}{\ln x}  \right|_{x = 0} \right),
\end{eqnarray}
where $V_{\rm A} = B/\sqrt{4 \pi n_{\rm i 0} m_{\rm i}}$ is the Alfv\'en
velocity and $\ocyci = eB/m_{\rm i}c$ is the cyclotron frequency of the
ions. In an usual accretion disk, $\Omega_{0}/\ocyci \ll 1$ holds and
also under the hybrid framework $V_{\rm A}/c \rightarrow 0$ is
assumed. Therefore the inequality of the charge due to the differential
rotation can be neglected in the hybrid framework. For the integration of
equation (\ref{eq_tidal_eom}) we use the ordinary Buneman-Boris method
with transformed electric/magnetic field,
\begin{eqnarray}
 \vec{E}^{\ast} &=& \vec{E} - 2 \frac{\mi}{e} q \Omega^{2}_{0} x
  \vec{e}_{x}, \label{eq_East} \\
 \vec{B}^{\ast} &=& \vec{B} + 2\frac{\mi c}{e} \vec{\Omega}_{0} . \label{eq_Bast}
\end{eqnarray}
For update of $\vec{E}$ and $\vec{B}$ we adopt the iterative method
proposed by \cite{Horowitz1989}.

\subsection{Initial Conditions and Simulation
  method}\label{subsec_init_cond}
As mentioned in the former section we investigate an evolution of the
system in the 2.5 dimensional meridional $(x,z)$ plane. The size of the
domain is set
to be $L_{x} \times L_{z} = 160 \times 960 $ cells whose grid interval
$\left( \Delta x \right)$ is set to be the half of the ion inertia length
$(\lini)$.
As an initial condition we adopt double Harris sheets
[\cite{Harris1962}] along $z$-direction, 
\begin{eqnarray}
B_{z0} \left( x \right) = B_{0} \left[  \tanh \left( \frac{x -
				      x_{\rm c}\left( t\right)}{l}
					      \right) - \tanh 
\left( \frac{x - x'_{\rm c}\left( t \right)}{l}  \right) - 1
				\right]. \label{eq_harrisB} 
\end{eqnarray}
Here $l$ is the width of the current sheet and is set to be
1.8$\lini$ throughout this article. $x_{\rm c}\left( t \right),x'_{\rm
c}\left( t \right)$ is the
center of each current sheet and as we shall discuss in Appendix 
\ref{appendix_cs}, the center of current sheet is not fixed in our
simulation model. Initially $x_{\rm c}(0)$ and $x'_{\rm c}(0)$ are
located on $L_{x}/4$, $3L_{x}/4$, respectively. To satisfy the initial
pressure
equilibrium ($B^{2}/8\pi = n_{\rm c}\left( T_{\rm i} + T_{\rm e}\right)$),
ions are assumed to have a spacial distribution of, 
\begin{eqnarray}
 n_{\rm i}(x)  = n_{\rm c} \left[ \cosh^{-2}\left( \frac{x - x_{\rm
					     c}\left( t\right)}{l} \right) +
 \cosh^{-2}\left( \frac{x - x'_{\rm
					     c}\left(t\right)}{l} \right) \right] +
 n_{0}, \label{eq_harrisN}
\end{eqnarray}
where $n_{\rm c}$ is the number density at the center of the current sheet
and $n_{0}$ is the number density in the outside of the current sheet
region. In the following
calculations we use 55 superparticles per cell in the outside of the
current sheet region. For the 
velocity distribution the ions are assumed to have a shifted-Maxwellian
distribution with a background Keplerian rotation,
\begin{eqnarray}
 f_{\rm i} \left( \vec{x}_{\rm i}, \vec{v}_{\rm i} \right) = n_{\rm
  i}\left( x \right)
  \left( \frac{\mi}{2 \pi T_{\rm i}} \right)^{3/2} \exp \left[ -
   \frac{\mi}{2 T_{\rm i}} \left( v_{x}^{2} + \left( v_{y} - v_{{\rm
						K}y} \left(
								   x\right)
					       - v_{\rm d,i}
					      \right)^{2}+ v_{z}^{2}
			   \right)\right], \label{eq_maxwellian} 
\end{eqnarray}
where $T_{\rm i}$ is the initial temperature of the ion and $v_{\rm
d,i}$ is the ion drift velocity which is defined as, 
\begin{eqnarray}
 v_{\rm d,i} = 2 \frac{T_{\rm i}}{B_{0} l}. \label{eq_Drift}
\end{eqnarray}
In the following, all physical quantities are normalized with those of
the parameters in the outside of the current sheet region. In the
 collisionless  accretion disks such as disks around a
black holes, the temperature of the electrons is considered to be much
colder than that of the ions [\cite{NarayanYiNature1995}]. Therefore we set
the initial temperature of the electron as $T_{\rm e} = 5\times10^{-3}
T_{\rm i}$. To save the 
integration time, we put localized perturbation in $y$-component of
the vector potential following the form introduced in \cite{Zenitani+2011},
\begin{eqnarray}
 \delta A_{y} = & &2l B_{1} \exp \left[ - \frac{\left( x - x_{\rm c}(t)
					   \right)^{2} + \left( z -
					   z_{\rm X} \right)^{2}}{(2l)^{2}
							 }\right]
 \nonumber   \\
 &-& 2l B_{1} \exp \left[ - \frac{\left( x - x'_{\rm c}(t)  \right)^{2}
		    + \left( 
			       z - z'_{\rm X} \right)^{2}}{(2l)^{2}} 
		 \right]  , \label{eq_PertB}
\end{eqnarray}
where $z_{\rm X} = L_{z}/4, z'_{\rm X}=3L_{z}/4$ is the $z$-coordinate
of the X-point and $B_{1}$ is the strength parameter of the perturbation
which is set to be $0.12 B_{0}$ in this calculation. In addition, we
include a localized resistivity in the X-points,
\begin{eqnarray}
 \eta = \eta_{0} &+& \eta_{\rm c} \cosh^{-2}\left[ \left( \frac{x -
						    x_{\rm c}(t)}{0.5
						    \lini} \right)^{2} +
					    \left( \frac{z - z_{\rm
					     X}}{\lini} \right)^{2} \right] \nonumber \\
 &+& \eta_{\rm c} \cosh^{-2}\left[ \left( \frac{x -
						    x'_{\rm c}(t)}{0.5
						    \lini} \right)^{2} +
					    \left( \frac{z - z'_{\rm
					     X}}{\lini} \right)^{2}
			    \right], \label{eq_resistivity}
\end{eqnarray}
and the magnetic Reynolds numbers defined with $\lambda_{\rm i} V_{\rm
A0} / \eta_{0}$ and $\lambda_{\rm i} V_{\rm A0} / \eta_{\rm c}$ are set
to be 2000 and 200, respectively. 
Here we impose the
uniform resistivity since the magnetic reconnection does not occur in
the ideal MHD limit. Therefore, we choose sufficiently small
value of the resistivity for the uniform component and compare our
results with the theory of the MRI under the ideal MHD approximation.
The localized resistivity is imposed as anomalous resistivity due to
the several instabilities generated in the vicinity of the X-point, such
as lower hybrid instability [e.g. \cite{HigashimoriHoshino2012}]. As we
will discuss in
Appendix \ref{appendix_cs}, we have confirmed that the following
results are not affected by the localized resistivity.
With this initial 
condition we perform five runs varying the rotational parameter $\left(
\Omega_{0}/\ocyci \right)$. We choose RUN A which has no rotation as a
fiducial run and gradually increase the rotational parameter. Several
important parameters in the calculation are described in
Table \ref{tab_parameter}. In RUN B and C the rotational parameter does
not exceed the instability criteria for the MRI defined with the parameters
in the outside of the current sheet region,
\begin{eqnarray}
 \Omega_{0}^{2} > \frac{k^{2} V_{\rm
  A,Out}^{2}}{3}. \label{eq_MRICriteria}
\end{eqnarray}
On the other hand with the rotational parameter used in RUN D and E,
the entire domain is unstable to the MRI. In RUN D resulting fastest
growing
wavelength of the MRI defined with the parameters in the outside of the
current sheet region is the same as $L_{z}$, whereas in RUN E the
fastest growing wavelength is the same as $L_{z}/2$. 

As we take the background differential rotation into account, the
evolution of the MRI is recovered in this hybrid code with a uniform
background plasma. In RUN F we impose uniform plasma to calculate a 
linear growth of the MRI. In high $\beta_{\rm i}$ plasma ($\beta_{\rm
i}\agt100$) rich kinetic effect of the ions such as generation of pressure
anisotropy [\cite{Hoshino2013}], and finite Larmor radius effect
[\cite{Ferraro2007}], would modify the linear behavior of MRI. However,
since we simply aim to focus on the conventional evolution of MRI, we set
$\beta_{\rm i}=1.0$ in this calculation. In
this run we impose 800 superparticles per cell whose grid interval was set
to be $\Delta x = \lini$. In this run the size of the domain
is set to be $L_{x}\times L_{z} = 80\times 480$ and the rotational
parameter is set to be the same as the one used in RUN E. Thus mode 2 of
the MRI is expected to be the fastest growing mode in this
calculation. The other parameters
 used in the simulation are listed in Table \ref{tab_parameter}. 
\begin{table}[htb]
 \caption{Simulation Parameters.}\label{tab_parameter}
  \begin{tabular}{c c c c c c c} \hline
 & RUN A & RUN B & RUN C& RUN D & RUN E & RUN F\\ \hline \hline
$L_{x}\times L_{z}$ & $160 \times 960 $ & $160
	   \times 960 $  & $160 \times 960 $  &
		   $160 \times 960 $  & $160 \times 960 $   & $80 \times
			   480$  \\
$\lambda_{\rm i}$ & 2$\Delta x$ & 2$\Delta x$ & 2$\Delta x$ & 2$\Delta x$ &
		       2$\Delta x$ &  $\Delta x$ \\
$n_{\rm c}/n_{0}$ & 4 & 4 & 4 & 4 &
		       4 & - \\
$\beta_{\rm i,Out}$ & 1.0 & 1.0 & 1.0 & 1.0 &
		       1.0 & $1.0$\footnote{As we consider uniform plasma, $\beta_{\rm i}=1.0$ is
 assumed in the entire domain in this run.} \\
$p_{\rm e}/p_{\rm i}$ & $5\times10^{-3}$ & $5\times10^{-3}$ &
	       $5\times10^{-3}$ & $5\times10^{-3}$ &
		       $5\times10^{-3}$ & $5.0\times10^{-3}$\\ 
$\Omega_{0}/\ocyci$ & 0 & $2\times 10^{-4}$ & $5\times 10^{-3}$ &
		   $1.35\times 10^{-2}$ & $2.70\times 10^{-2}$ &
			   $2.70\times 10^{-2}$ \\ 
Current Sheet ? & YES & YES & YES & YES& YES & NO \\\hline
  \end{tabular}
\end{table}



\section{Results}\label{sec_results}
\subsection{Linear growth of MRI}\label{subsec_mri}

\begin{figure}[htbp]
  \begin{center}
   \includegraphics[bb=15 30 355 245,
   width=\hsize, angle=0]{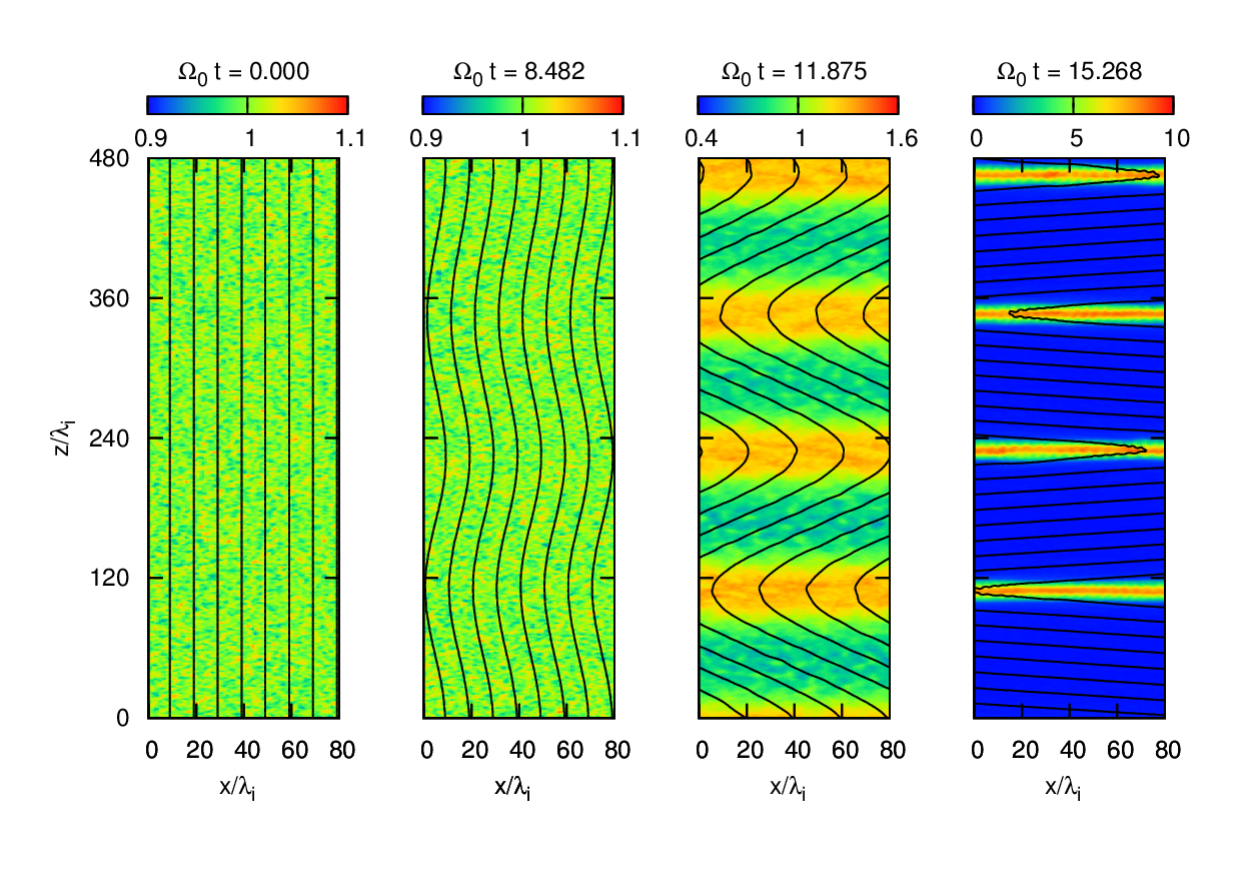}
  \end{center}  
 \caption{Time evolution of the MRI. The color contour corresponds to the ion
 density and the solid line corresponds to the magnetic field line.}
 \label{fig_MRIEvolution}
\end{figure}

First, we briefly look at the evolution of
the MRI. Figure \ref{fig_MRIEvolution}
shows the time evolution of the density and the magnetic field line in
RUN F. As expected in the previous section, mode 2 of the MRI 
which can be clearly observed in Figure \ref{fig_MRIEvolution}, is
the most unstable with the parameters used here. 
As it is often reported in the MHD calculations, two sets of channel flows
are generated in the final state of the calculation.

\begin{figure}[htbp]
   \begin{center}
    \includegraphics[bb=10 30 700 500,
    width=\hsize, angle=0]{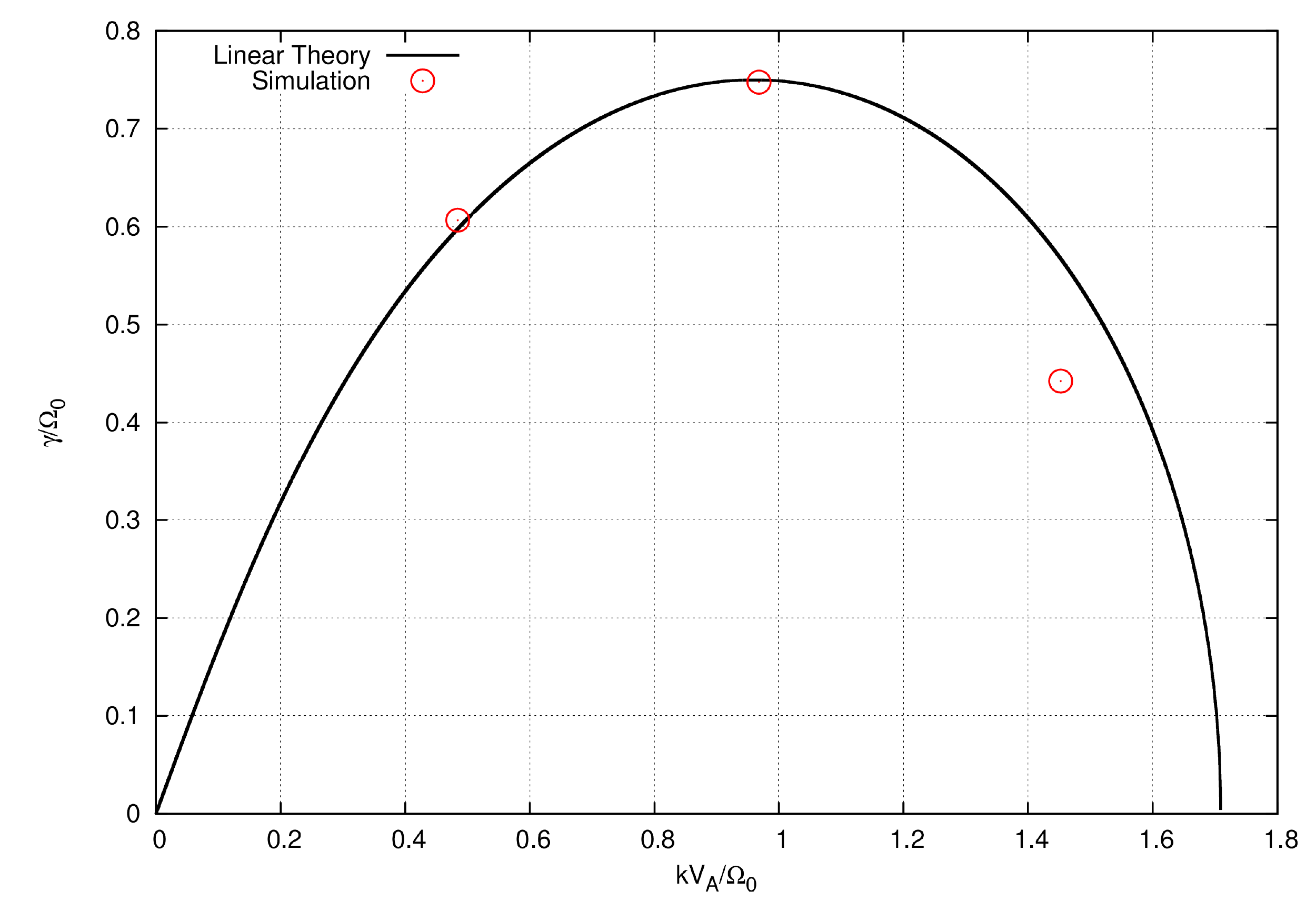}
   \end{center}  
 \caption{The dispersion relation of the MRI. The open circles
 corresponds to a
 growth rate obtained from the simulation. Error bars are smaller than
 the size of the open circles.}
 \label{fig_MRIGrowth}
\end{figure}

The linear growth of the MRI is also confirmed with the Fourier
decomposition. In Figure \ref{fig_MRIGrowth} we show a growth rate obtained
from our simulation. The horizontal axis corresponds to the normalized
wavenumber whereas the vertical axis corresponds to the growth
rate. Solid curve corresponds to a dispersion relation of the MRI obtained
from the linear analysis of a Hall MHD equations
[\cite{BalbusTerquem2001}], whereas the open circles correspond to the
results from our hybrid simulation.
 As we can see from the plot, our results are in good
agreement with those of the conventional MRI.

\subsection{Magnetic Reconnection\cl Overview}\label{subsec_overview}
Let us go on to the results of the magnetic reconnection. We first overview
our results. Figure \ref{fig_denline} shows the 
structures of five runs at the each stage of the reconnection. The most
distinct feature is found in RUN E whose magnetic field line in the
entire region of the simulation domain is remarkably bent. We consider
that in this parameter the MRI
is the dominant process for the evolution of the system. The relation
between the magnetic reconnection and the MRI shall be discussed in the
following section.
\begin{figure}[htbp]

  \begin{center}
   \includegraphics[bb=10 20 364 254,
   width=0.92\hsize, angle=0]{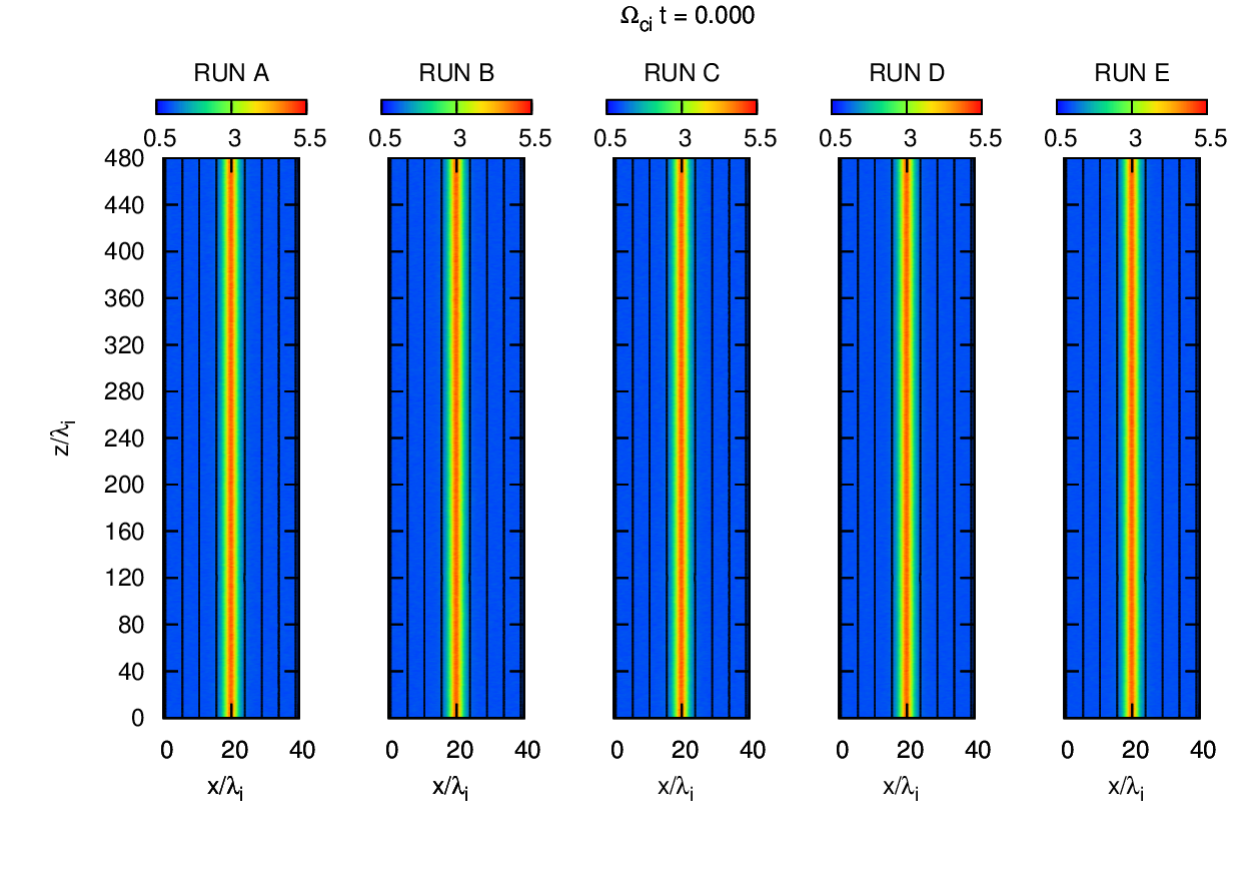}
  \end{center}  

  \begin{center}
   \includegraphics[bb=10 40 364 254,
   width=0.92\hsize, angle=0]{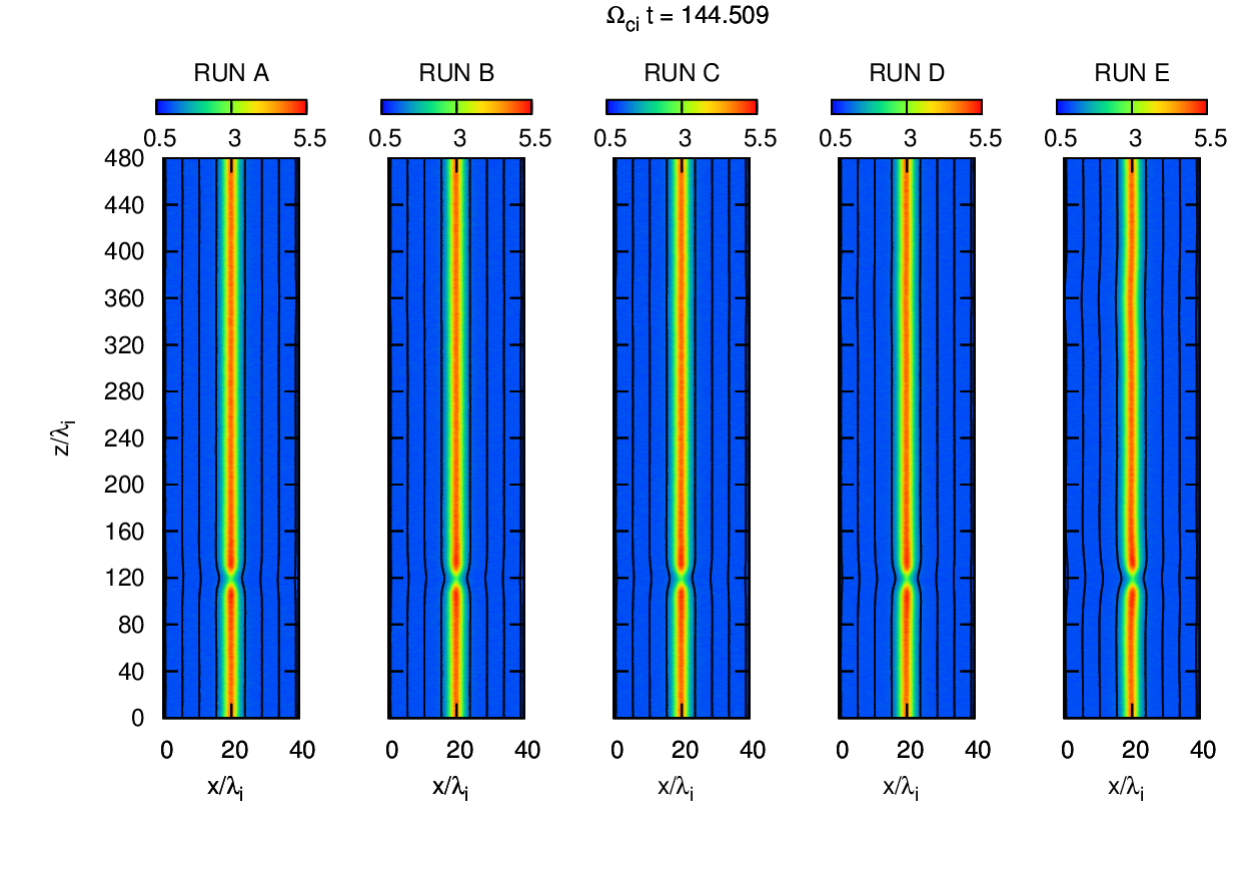}
  \end{center}  

\caption{The time evolution of all runs. The color contour corresponds
 to ion
 density whereas the solid line corresponds to the magnetic field line. The
 figures are focused on the left half of the simulation domain.}
 \label{fig_denline}
\end{figure}
\addtocounter{figure}{-1}
\begin{figure}[htbp]
  \begin{center}
   \includegraphics[bb=10 20 364 254,
   width=0.92\hsize, angle=0]{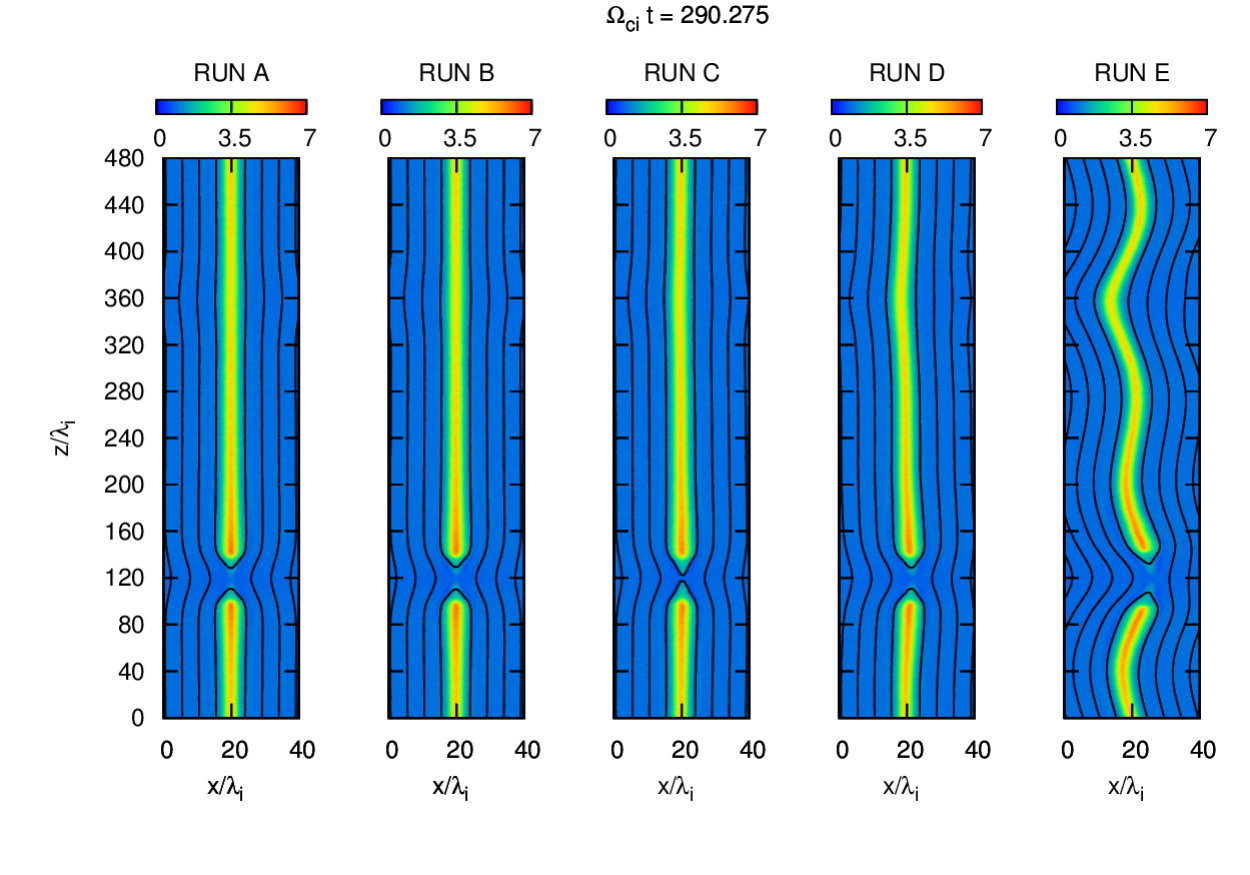}
  \end{center}  

  \begin{center}
   \includegraphics[bb=10 40 364 254,
   width=0.92\hsize, angle=0]{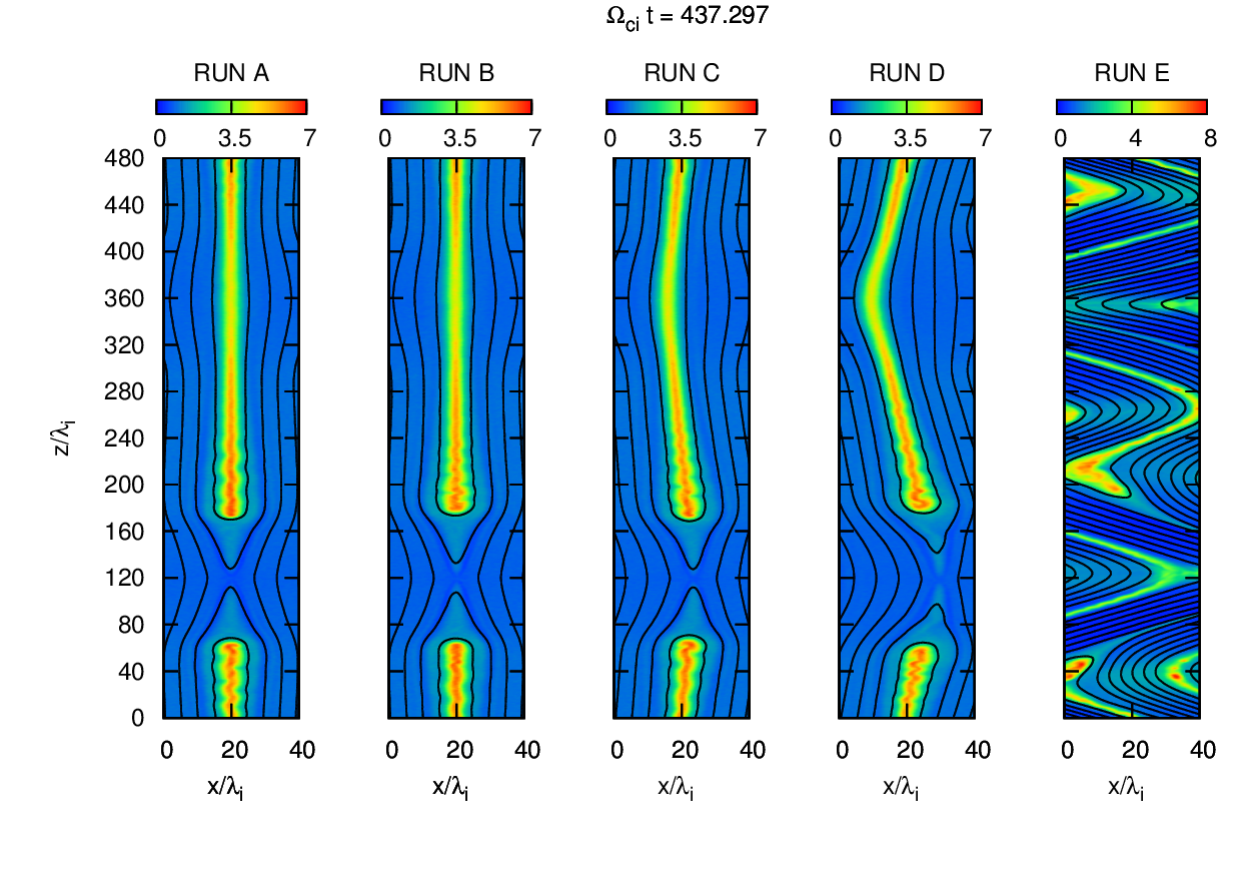}
  \end{center}  
\caption{Continued.}
\end{figure}
In RUN B,C and D the effect of the Kepler rotation seems to be relatively
moderate compared to RUN E. At the onset stage of the reconnection ($\ocyci
t\simeq$144) there is no significant difference in the structure between
each run. However, as the reconnection goes on, an asymmetric structure
gradually becomes remarkable in RUN B, C and D. The asymmetry becomes
clearer as the rotational parameter increases. In addition, a
migration of X-point is observed in all run with the finite Kepler rotation
(RUN B-E). A physical interpretation of this migration shall also be
discussed in the following section.
Another asymmetry is also found in the out-of-plane magnetic
field. Figure \ref{fig_HallList}
shows the structure of the out-of-plane magnetic field ($B_{y}$) at the
final stage of five runs. It is clear,
especially in RUN C and D, the absolute value of $B_{y}$ is, in average,
larger in the right side of the X-point compared to that in the left
side. This is understood by considering the coupling of the Hall effect and
 the differential rotation and shall be discussed in the next section.

\begin{figure}[htbp]
\begin{center}
 \includegraphics[bb=20 40 426 300,
  width=\hsize]{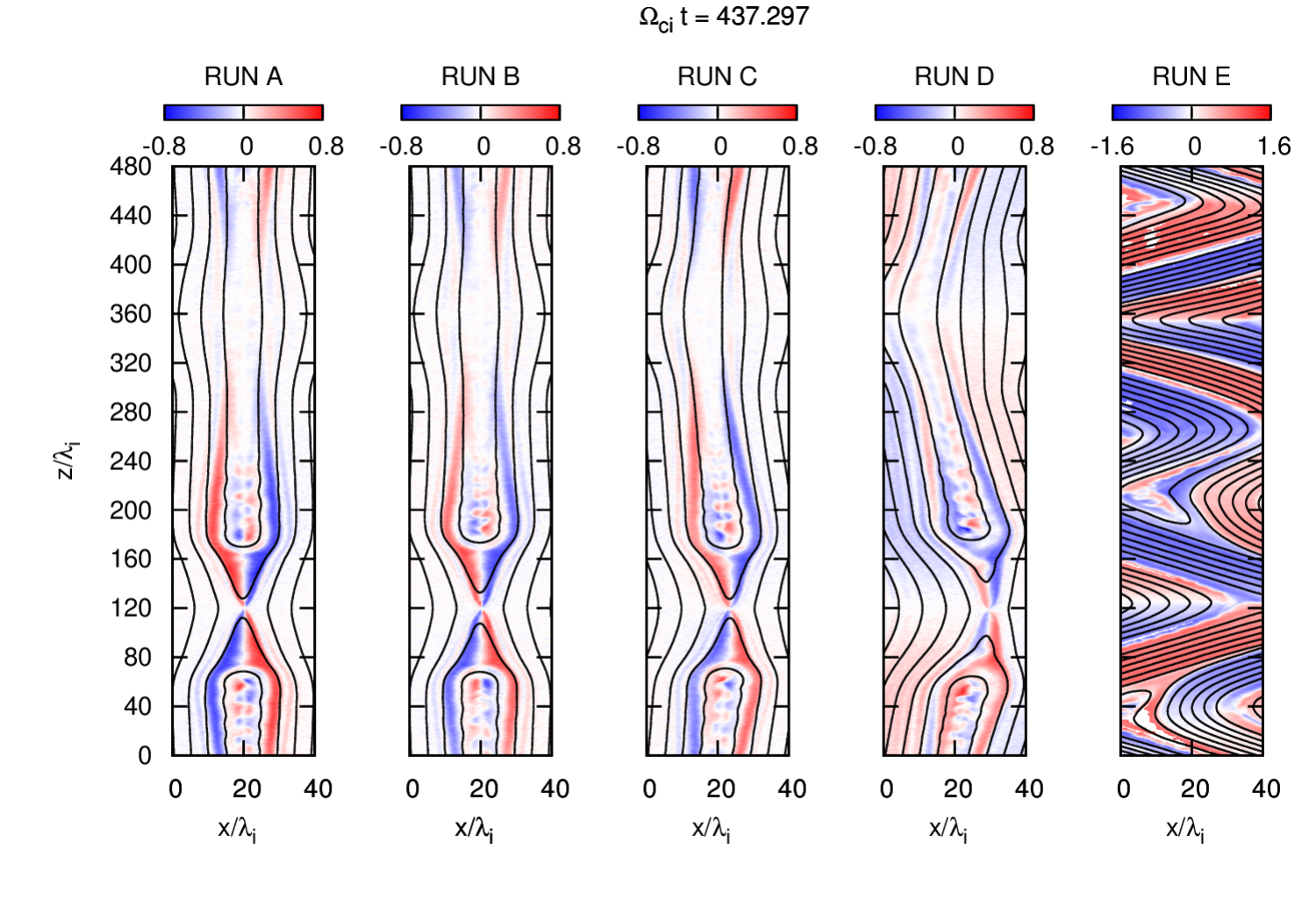}
\end{center}
\caption{The out-of-plane magnetic field ($B_{y}$) in the final stage of
 the
 simulation. $B_{y}$ in RUN E is remarkably enhanced due to the dynamo
 effect of the MRI. The figures are focused on the left half of the
 simulation domain.}
 \label{fig_HallList}
\end{figure}

\subsection{Magnetic Reconnection\cl Asymmetric evolution of
  out-of-plane magnetic field}\label{subsec_hallfield}
It is well known that the quadrupole structure in the out-of-plane
magnetic field is observed around the X-point in the numerical
simulation under 
the framework which distinguishes the magnetization degree of the ions and
the electrons.
In the hybrid framework which treats the ions as superparticles with finite
mass and the electrons as massless charge neutralizing fluid, the
magnetization
feature of each component is also distinguished.
This gives a significant difference in the motion of each component around 
the X-point resulting the in-plane Hall current and out-of-plane
magnetic field
[\cite{Sonnerup1979Text}]. 

\begin{figure}[htbp]
 \begin{center}
 \includegraphics[bb=20 30 370 248,
  width=\hsize]{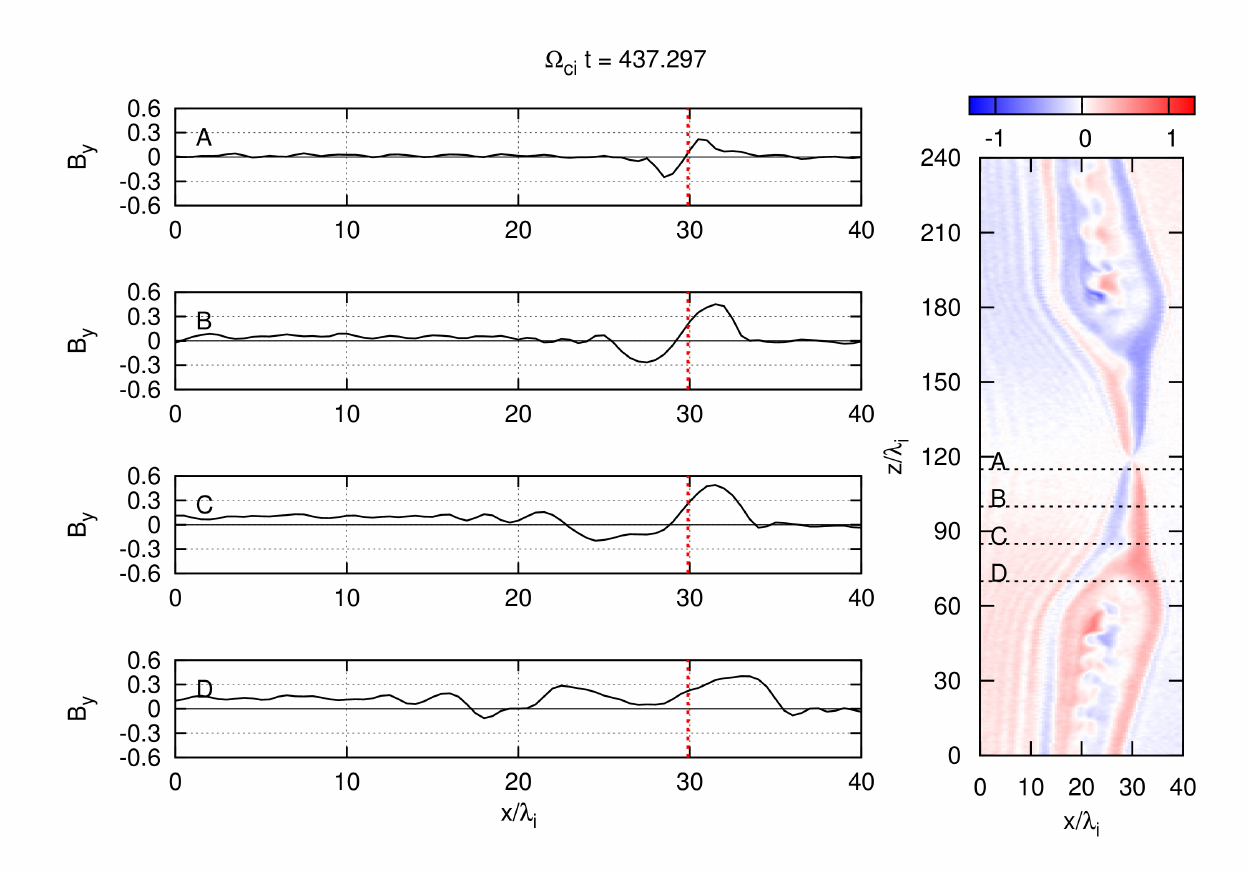}
 \end{center}
 \caption{Focused view of out-of-plane magnetic field in RUN D.}
 \label{fig_Hall_F}
\end{figure}

In our fiducial run (RUN A),
a clear quadrupole structure is also found in the symmetric manner
(Figure \ref{fig_HallList}). As we pointed out in the previous section
an asymmetric structure in the
out-of-plane magnetic field ($B_{y}$) is found in RUN B, C and D. The
asymmetry becomes significant as the rotational parameter increases.
 Figure \ref{fig_Hall_F} shows a focused
view of the out-of-plane magnetic field ($B_{y}$) at $\ocyci t \simeq 437$
of RUN D. The color contour in the right panel corresponds to
$B_{y}$. In the left panel of Figure \ref{fig_Hall_F}, $B_{y}$ is also
plotted along the dotted line denoted in the right panel, and from the
top to the bottom each panel corresponds to line A-D. The dashed red lines
plotted vertically in the left panels correspond to the baseline which
passes through the X-point. In cut C and D clear asymmetry with
respect to the red line is observed and the absolute value of $B_{y}$ is
always found to be large in the right side of the X-point compared to
the left side. These structures can be understood by a simple coupling
between the Hall magnetic field produced by the reconnection and the
effect of the differential rotation.

\begin{figure}[htbp]
 \begin{center}
 \includegraphics[bb=20 20 1000 760,
  width=\hsize]{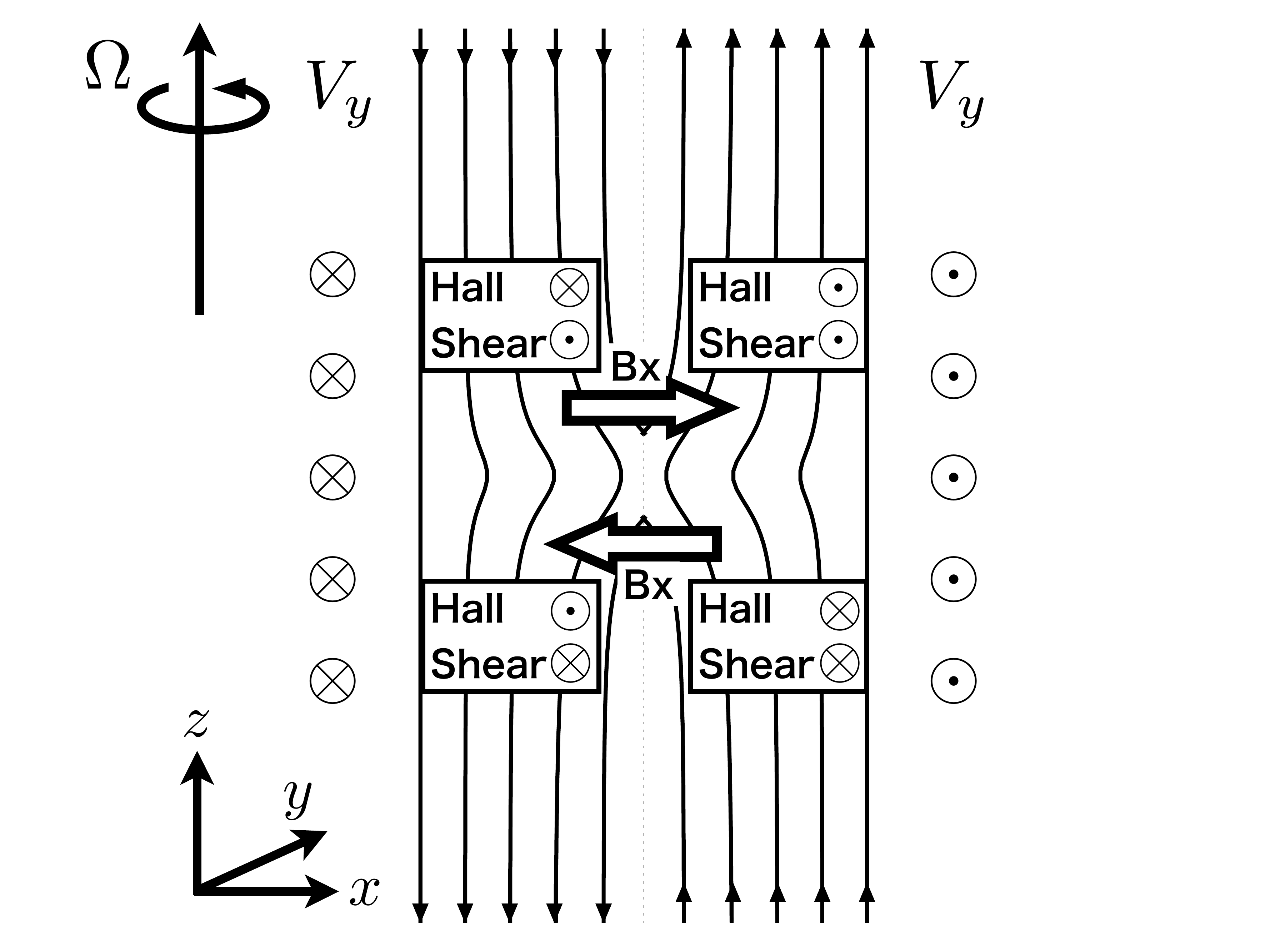}
 \end{center}
 \caption{A schematic plot of the Hall Field and the sheared Field.}
 \label{fig_Hall_Sch}
\end{figure}

Figure \ref{fig_Hall_Sch} shows the schematic view of the system. 
As the reconnection grows the quadrupole magnetic field parallel to the
$y$-axis is generated due to the Hall effect. As the reconnection goes
on and $B_{x}$ is generated the dynamo effect of the differential rotation
produces $B_{y}$ following to the $y$-component of the induction
equation (\ref{eq_ByShear}). Since $B_{x}$ generated by the reconnection
has opposite sign on the different side of the X-point so as $B_{y}$
generated by the shear motion. Superposing the quadrupole $B_{y}$ due to
the Hall current and the antisymmetric $B_{y}$ due to the shear motion,
the asymmetric structure in the out-of-plane magnetic field is clearly
understood and in the case of Figure \ref{fig_Hall_Sch}, for instance, the
larger absolute value of $B_{y}$ is observed in the right side of the
X-point.

Paying attention to the fine structure, the ``undulated'' feature in
the out-of-plane magnetic field is
found in the nonlinear stage of the reconnection. This structure is
consistent with the one introduced in \cite{ArznerScholerJGR2001}. They
have pointed out that the structure is a consequence of an instability
driven by a shear flow and a parallel pressure anisotropy generated
in the layer consisted with the outflow region and the outside of the
current sheet region. We note that in
our simulation setup with the finite rotational parameters, any
scale of perturbation would be unstable to the MRI inside the current
sheet following to the conventional definition of the MRI criteria. This
is because inside the current sheet the magnetic field
gradually decreases whereas the density increases which leads to $V_{\rm
A,Out}\rightarrow 0$ in the right hand side of the equation
(\ref{eq_MRICriteria}). However as we can see from
Figure \ref{fig_HallList}, the
characteristic scale of the ``undulated'' structure seems to be
insensitive to the rotational parameter. Thus this structure can also
be considered as the one introduced in \cite{ArznerScholerJGR2001}
and the modulation by the MRI seems to be rather weak in this instability.
In addition, a standing wave structure whose wave front is approximately
parallel to the magnetic field line is found in the region where the
transition takes place from the current sheet to the
outside region. Approximating the structure of
current sheet as a slow shock, this wave structure is
regarded as a standing wave train found in the dispersive shock when the
ion inertia scale is much larger than the resistive scale
[\cite{HauSonnerup1992}]. These fine structures are found not only in
RUN D but also in RUN A-C on which the effect of the Kepler rotation is
relatively moderate.

\subsection{Magnetic Reconnection\cl Migration of
  the X-points}\label{subsec_xpointmig}
\begin{figure}[htbp]
 \begin{center}
 \includegraphics[bb=20 30 370 250,
  width=\hsize]{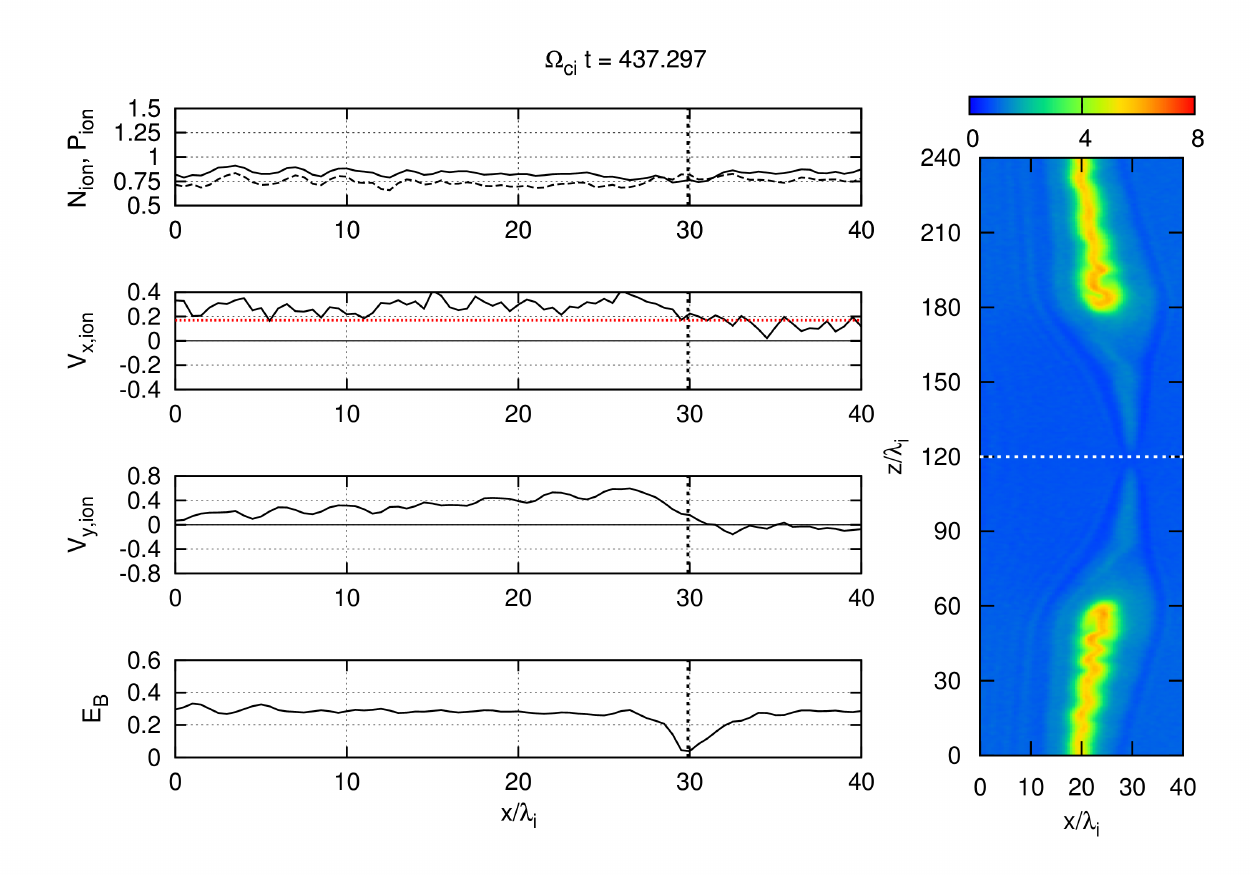}
 \end{center}
 \caption{A focused view of the X-point in RUN D.}
 \label{fig_XP_mig}
\end{figure}

Another remarkable structure is found in the vicinity of the
X-point. As described in the overview the asymmetry becomes
significant as time goes by. At the same time, the X-point migrates in a
certain direction. 
The direction of the X-point migration is always the same as long as the
sign of $\vec{J}_{0}\vec{\times}\vec{\Omega}_{0}$, and we observe the
migration in the opposite direction with the opposite sign of
$\vec{J}_{0}\vec{\times}\vec{\Omega}_{0}$,.
The migration distance of the X-point is larger in
the runs with the larger rotational parameter.
Figure \ref{fig_XP_mig} shows a focused view of the X-point at 
$\ocyci t \simeq 437$ in RUN D. Here, the color contour in the right panel
corresponds to the ion density. In the left panels, we show several
physical values along the dashed white line indicated in the right panel.
From the top, the density (solid line) and the pressure (dashed line) of
the ions, the moment averaged ion velocity ($V_{x,{\rm ions}}$), the
$y$-component ion velocity deviation from the background Kepler rotation
($V_{y,{\rm ions}}$), and the
magnetic energy normalized with the initial value in the outside of the
current sheet region, are plotted.
The horizontal red dashed line plotted in the second row 
corresponds to the migration velocity of the X-point. In average,
inflow speed of the ions from the both sides of the X-point are the same
in the rest frame of the X-point.

The migration of the X-point itself is easily understood by the
generation of the flow by the MRI. However the correlation between the
sign of $\vec{J}_{0}\vec{\times}\vec{\Omega}_{0}$ and the direction of
the migration is not obvious.
This correlation can be understood by considering the asymmetric growth
of MRI with respect to the neutral sheet. 
Since the conventional criteria of the MRI
is given by Equation (\ref{eq_MRICriteria}), the MRI is always active
inside the current sheet where the magnetic field strength and the
ambient Alfv\'en velocity are infinitesimally small. However, since we
have carried out the calculations under hybrid framework, the effect of
the Hall term should also be taken into account. As reported in
\cite{BalbusTerquem2001}, the Hall term extends the unstable region of
the MRI to a shorter wavelength when the orientation of the magnetic
field and the background rotational vector is anti-parallel, and vice
versa. Following to their dispersion relation, the effect of Hall term
is still minor in the outside of the current sheet region in our
simulations. However, in the vicinity of the current sheet, the time
scale of the cyclotron motion may reach to that of the Kepler rotation,
and as a result the Hall term would be effective in this region. In
Figure \ref{fig_XP_mig}, the
ambient magnetic field is anti-parallel to the rotational vector in the
left side of the current sheet and the instability criteria is reduced
in this region. Therefore the small scale perturbation triggered by the
magnetic reconnection would couple to the MRI and can be enhanced in
this side of the current sheet, resulting
a migration of the X-point, whereas in the opposite side of the current
sheet, the characteristic scale imposed by the magnetic reconnection
does not satisfy the criteria of the MRI under the Hall effect. From the
second and the third row, we find $V_{x, {\rm ions}}/V_{y, {\rm
ions}}>0$ in the left side of the X-point which is also satisfied during
the growth stage of the MRI. We consider this phase relation is also
consistent with the above explanation.
 In addition to the calculation introduced in the present article, we
 have also varied the number of the superparticles and the grid interval
 and confirmed that the above results were not affected by these
 parameters.

\subsection{Magnetic Reconnection\cl\ Magnetic reconnection as a trigger
  of the MRI}\label{subsec_growth}
As we have described in the overview, in the outside of the current sheet
region of RUN E
perpendicular magnetic field is remarkably enhanced which implies the
dominant process in this parameter regime is the MRI. During the evolution
of the MRI, a magnetic reconnection also takes place with the
associating X-point migration giving additional growth of $B_{x}$
together with the reconnection. In this section,
we point out that the reconnection would also 
contribute to a relatively large initial perturbation on the growth of
the MRI.

\begin{figure}[htbp]
\begin{minipage}{1.0\hsize}
\begin{center}
 \includegraphics[bb=10 20 360 250,
 width=\hsize, angle=0]{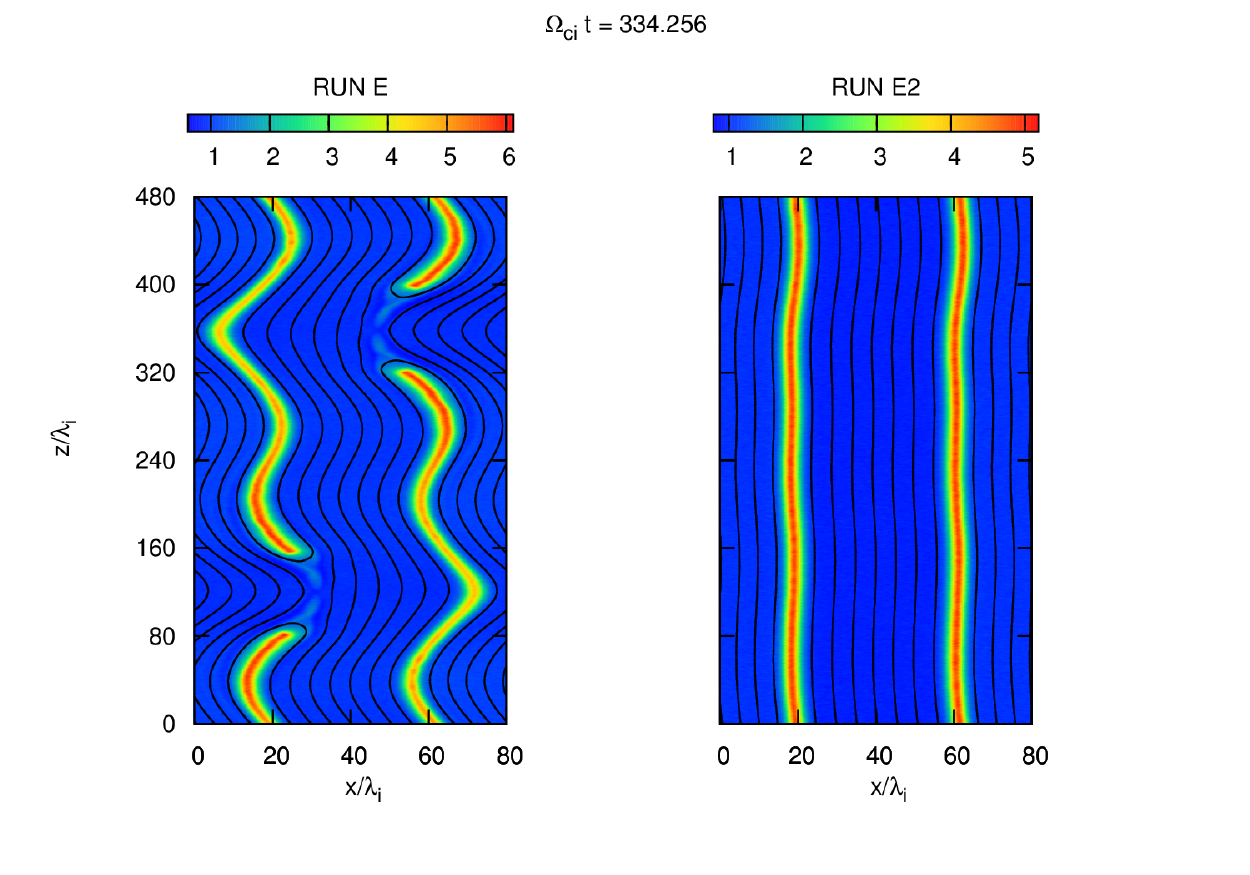}
\end{center}
\end{minipage}\\
\begin{minipage}{1.0\hsize}
\begin{center}
 \includegraphics[bb=10 20 360 250,
  width=\hsize, angle=0]{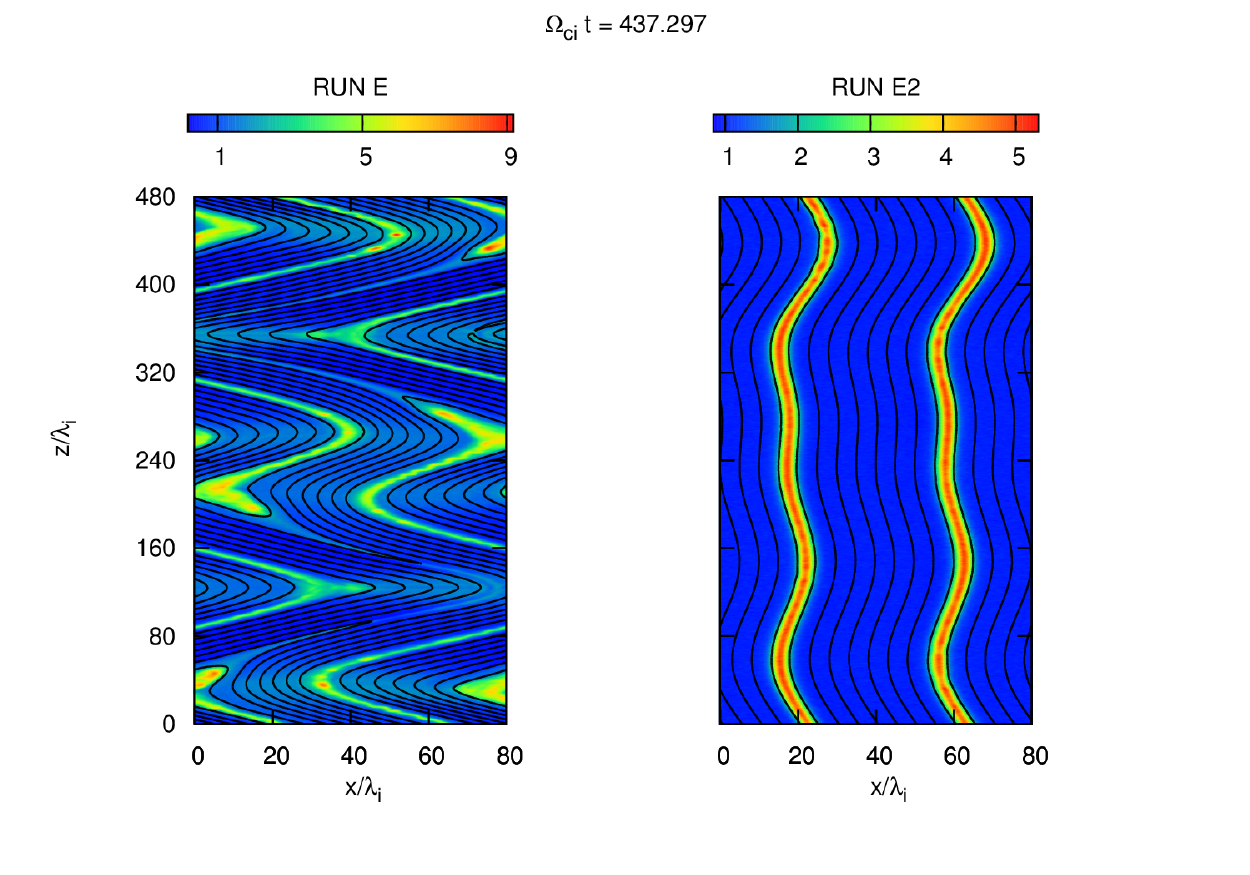}
\end{center}
\end{minipage}
\caption{The density (color) and the magnetic field line (solid line) at
 the middle stage of RUN E and RUN E2 (top), and at the final stage
 (bottom).}
 \label{fig_CompMRI}
\end{figure}

In order to evaluate the effect of the magnetic reconnection on the
evolution of the MRI we perform another comparative calculation.
Here we choose RUN E as a fiducial run and another
 run (RUN E2) is calculated without the initial trigger
for the magnetic reconnection$\scl$ by excluding the initial vector
potential (\ref{eq_PertB}) and taking $\eta_{\rm c} = 0$ in the resistivity
(\ref{eq_resistivity}). Note that we have left the uniform component of
the resistivity
$\eta_{0}$. Therefore, under this initial condition, both a tearing mode
instability and an MRI would take place. Here the fastest growing mode of
the MRI in the outside of the current sheet region is 0.75$\Omega_{0}
\simeq 0.02\ocyci$ which is comparable
to that of a tearing mode which can roughly be estimated as
$10^{-1}$-$10^{-2}\ocyci$
[e.g. \cite{Terasawa1983}\scl\cite{HesseWinske1993GRL}]. 

Figure \ref{fig_CompMRI} shows the time evolution of the two comparative
runs. Again the color contour corresponds to the ion density, and the
solid line corresponds to the magnetic field line. From the top panel
($\ocyci t \simeq 334$) we observe that the X-point in the different
current sheet migrates in the opposite direction. From the bottom panel
($\ocyci t \simeq 437$) it is clearly observed that the perpendicular
component of the magnetic field is significantly enhanced in RUN E.
\begin{figure}[htbp]
 \begin{center}
 \includegraphics[bb=20 70 1000 760,
  width=\hsize]{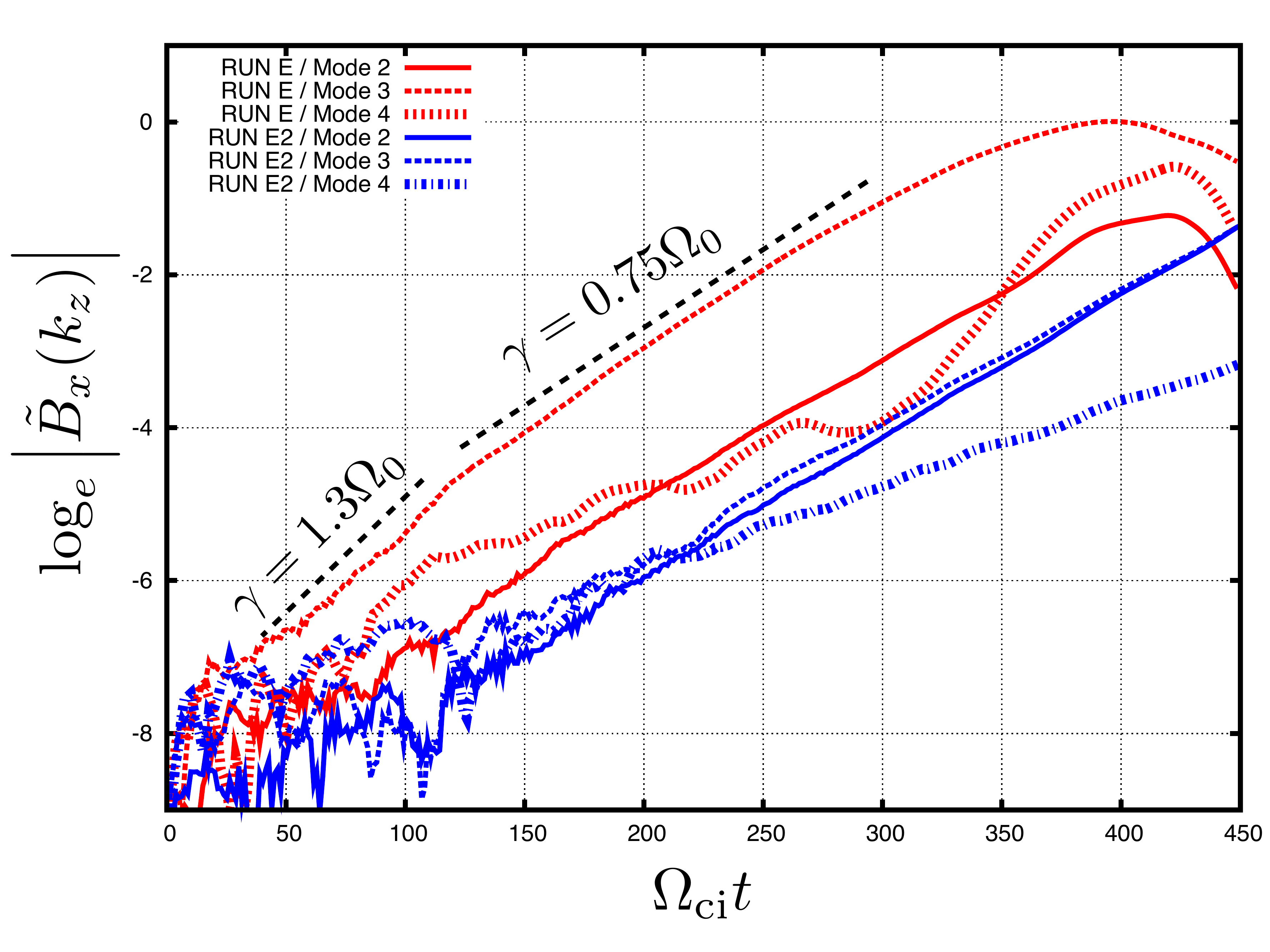}
 \end{center}
 \caption{The time evolution of Magnetic Field Power.}
 \label{fig_Growth}
\end{figure}\\
Another distinct feature is also found from the Fourier decomposition of
 $\left| B_{x}\right|$ in each run. In Figure \ref{fig_Growth} we show
 the evolution curves of each mode in the both calculations.
The Fourier transformation was applied on
$B_{x}$ along the $z$-axis in the outside of the current sheet region
 ($10\lambda_{\rm i} \le x \le  20 \lini$). In this
region, we can approximate the uniform plasma background and can compare
results with the theory of conventional MRI. Within the two runs we find no
significant difference in the maximum growth rate and  was the same
as that of the conventional MRI $\left( \gamma_{\rm
max}=0.75
\Omega_{0} \right)$ except for mode 3 in RUN E. During the
initial growth
stage of RUN E (0$\ocyci t-$100$\ocyci t$) the growth rate is roughly
1.3$\Omega_{0}$ which is about 1.8 times faster than that of the MRI. This
implies
that the magnetic reconnection and the associated X-point migration
 initially
gives relatively large perturbation and act as an trigger for the 
MRI. During the middle stage of the evolution, the growth rates of all
the modes do not exceed the maximum growth rate of the MRI. Triggered by
 the X-point migration, in this
stage, the power of $B_{x}$ in the outside of the current sheet  region of
RUN E is, in average, $\sim30$ times larger than that of RUN E2. 
It is worth noting that the fastest growing mode in RUN E is mode 3 even
the initial rotational parameter was set for mode 2 to be the most
 unstable. At the initial stage of reconnection, the reconnection rate
corresponding to the separatrix angle determines the mode with the
largest amplitude. At the same time, the X-point migration from the another
current sheet gives perturbation towards opposite direction in $z \simeq
3L_{z}/4$. As a result at the initial growth stage a large amplitude
perturbation corresponding to mode 3 is given by the magnetic
reconnection and by the associating X-point migration. During the middle
 stage of the growth ($\ocyci t \agt 100$) the ambient
magnetic field parallel to the rotational axis is weakened by magnetic
reconnection leading shorter wavelength to be more unstable to the
 MRI. It is well known that the linear process of a 2.5D MRI
also holds even the perturbed field amplitude is relatively large
compared to the ambient field [\cite{GoodmanXu1994}]. Therefore, in the
middle stage of the evolution, mode 3 can be the most unstable.



\section{Summary and Discussions}\label{sec_dis}
In this study we developed a 2.5D hybrid code including the Coriolis and
the tidal force which can be found in a local corotating frame with a
standard open shearing boundary condition [\cite{Hawley+1995}]. In this
code, azimuthal symmetry is assumed and as a result, the boundary
condition degenerates to a standard periodic boundary condition for
the deviation components. With this code we first
confirmed the evolution of the MRI. As a first step we focused on the
linear growth of the instability in the low $\beta_{\rm i}$ limit so
that we would neglect rich kinetic effects introduced in
\cite{Hoshino2013} and \cite{Ferraro2007} and would compare the results
with the conventional linear theory. The result was in good
agreement with the conventional theory of the MRI. We note that the
studies of long term evolution of MRI in the high $\beta_{\rm i}$ regime
would be important to understand the relaxation process of the pressure
anisotropy in the collisionless accretion disks.
%

We also investigated the effect of the differential rotation on the
evolution of the magnetic reconnection with five consecutive runs.
Main results are summarized as follows\cl
 
First we found an asymmetric structure in the out-of-plane magnetic
field in the vicinity of the X-point. This can be understood by the
coupling
of the Hall term and the differential rotation. As the reconnection grows,
 a Hall current around the X-point
generates quadrupole magnetic field. At the same time, the perpendicular
magnetic field generated by the magnetic reconnection is sheared by the
differential rotation generating a dipole structure in the out-of-plane
magnetic field. The asymmetry is created as a superposition of these two
processes. Furthermore, we also confirmed the ``undulated'' structure
reported in \cite{ArznerScholerJGR2001}. Here, the differential rotation
seems to have a minor effect on the generation of the ``undulated''
structure.

Next we found a migration of the X-point. This can be interpreted as
a result of the MRI evolution coupled with the magnetic reconnection
under the Hall effect. Since we have taken the Hall effect into account,
the instability criteria of the MRI is different between the opposite
sides of the neutral sheet.
The particular side of the outside of the current sheet region has a
reduced criteria of the MRI, whereas the another one has a severer
criteria. Since a small scale perturbation given by the magnetic
reconnection would couple with the MRI under the Hall effect in the
particular side of the current sheet, the
perturbation imposed by the magnetic reconnection would be enhanced in
an asymmetric manner. This asymmetry in a growth of the MRI would result
the migration of the X-point and its direction is determined only by the
initial sign of $\vec{J}_{0} \vec{\times \Omega}_{0}$. 


Finally we found that the migration of the X-point can be a trigger of
the MRI and can modify the characteristic scale of the perturbation. As
the X-point migration generates relatively large $B_{x}$ perturbation,
the magnetic reconnection in the differentially rotating system can be
an effective trigger of the MRI. At the same time, the magnetic
reconnection annihilates the ambient magnetic field shifting the most
unstable wavelength of the MRI to the smaller scale.

It is well known that during the quasi steady state of the MRI induced
turbulence, the magnetic energy is enhanced intermittently and released
within a short time scale. During this stage a disruption and
a re-organization of the channel flow take place repeatedly and a
magnetic reconnection plays an important role on the dissipation of the
magnetic field and heating of plasma.
This process generates spike structures in the time evolution of the 
Maxwell stress curve during the quasi steady state
[\cite{SanoInutsuka2001}]. It has also been pointed out from 2.5D PIC
simulations that during the ``active'' phase when the magnetic
reconnection takes place repeatedly the efficiency of the angular momentum
transport increases [\cite{Hoshino2013}]. These results imply that the 
repeating process of magnetic energy enhancement by the MRI and
dissipation by the magnetic reconnection is essential to an effective
angular momentum transport. 
Though our initial setting is limited in the specific structure of the
current sheet, we consider that the ``triggering'' effect of magnetic
reconnection would be an effective seed of a re-organization of the current
sheet associated to the channel flow and as a result, contributes to a
strong angular momentum transport in the collisionless accretion disk.

It is worth noting by using the hybrid code, we have bridged the scale
gap between the MHD scale and the PIC scale. Though several hundreds of ion
inertia scale is still be smaller than the actual scale of the disk, we
believe that the results of the present article would be non-negligible in
the thick current sheet. For fully understanding the evolution of the MRI,
the associated generation and relaxation of pressure anisotropy,
the dissipation by magnetic reconnection, and the resulting efficient
angular momentum transport, a massive 3D simulation would be
required. We would
also like to point out that not only the 2.5D meridional plane analysis
but also the equatorial ($x,y$) plane analysis would be important for
the understanding of the fundamental physics of the magnetic
reconnection in the differentially rotating system.
\\

The author would like to acknowledge T.Amano, K.Higashimori, and
K.Hirabayashi for valuable and insightful discussions. This research is
supported in part by Grant-in-Aid for Japan Society for the Promotion of
Science (JSPS) Fellows (No.10J08225), and JSPS Kakenhi (No.25287151).


\appendix
\section{\label{appendix_sh_bnd} Methods for calculating differentially
 rotating system using hybrid code.} 
Here we briefly describe about the modification applied in our hybrid code
due to the differential rotation. 

As we explained in Sec. \ref{sec_simulation}, we assume that the
background Kepler rotation always exists. Therefore
one must pay attention to the boundary condition since the out-of-plane
flow velocity differs between the inner and outer boundary of the domain. 
However, as a Keplerian velocity can be exactly calculated from
equation (\ref{eq_vkepler}) we simply focus on the deviation from the
Keplerian profile.
Since we assume the azimuthal symmetry ($\partial / \partial y=0$) the open
shearing boundary condition proposed by \cite{Hawley+1995} degenerates to a
standard periodic boundary condition for the deviation components.

For updating of the particles, we use the Buneman-Boris method. Here we use
the particle velocity
observed in the corotating frame\scl that is, all the particle velocity
includes the Keplerian component in the time integration of the
particle. At each time step we subtract the Keplerian velocity component
of the particles using updated position information $( \vec{x}_{\rm
i}^{n+1/2})$. Then we apply a moment calculation to
obtain the deviation component of the current density using an appropriate
shape function,
\begin{eqnarray}
 \vec{J}^{n+1/2}_{\rm i,dev} = e \sum_{\rm particle} \vec{v}_{\rm
  i,dev}^{n+1/2}
  S(\vec{x}_{\rm i}^{n+1/2}). \label{eq_shape}
\end{eqnarray}
From $\vec{J}_{\rm i,dev}^{n+1/2}$ we calculate the deviation component of
the electric field $( \vec{E}_{\rm dev}^{n+1/2})$ by using the Ohm's law
(\ref{eq_Ohm}), and the Amp\'ere's law (\ref{eq_Ampere}). Note that since
the simulation is performed under
nonrelativistic limit, the magnetic field is Lorentz invariant. The
electric field observed in the corotating frame is obtained by applying
a Lorentz transformation on this electric field. This is applied by
adding the induction term due to the Keplerian flow to the deviation
component,
\begin{eqnarray}
 \vec{E}^{n+1/2} = \vec{E}_{\rm dev}^{n+1/2} - \frac{1}{c} \vec{v}_{\rm
  K} \times \vec{B}^{n+1/2}. \label{eq_Eadd}
\end{eqnarray}
For the update of the magnetic field, we separate the $\vec{\nabla \times}$
calculation of the electric field to the deviation component and the
Keplerian
component. As we assume the azimuthal symmetry ($\partial/\partial y=0$),
together with the solenoidal condition (\ref{eq_solenoidal}), the
Keplerian component of the induction term is calculated as follows,
\begin{eqnarray}
 \frac{1}{c} \vec{\nabla \times} \left( \vec{v}_{\rm K} \times
				  \vec{B}^{n+1/2}
				 \right) = - \frac{1}{c} q \Omega_{0}
 B_{x}^{n+1/2} \vec{e}_{y}. \label{eq_ByShear}
\end{eqnarray}
This equation
physically describes the shearing effect of the differential
rotation generating the $y$-component magnetic field from the
$x$-component.
$\vec{B}^{n+1/2}$ and $\vec{E}^{n+1/2}$ are iteratively calculated in the
update of the field. Since the periodicity is guaranteed for the deviation
components and for the magnetic field, this separation of the $\vec{\nabla
\times}$ calculation greatly reduces the
complication in taking the boundary condition with the shear flow.

\section{\label{appendix_cs} Harris solution in differentially rotating
system} 
Here we discuss the behavior of a Harris solution in the rotating system.
A conventional Harris solution cannot be an exact kinetic equilibrium
solution in a differentially rotating disk
because the solution introduced in \cite{Harris1962} is based on
the conservation of the canonical angular momentum for each particle.
In a differentially rotating disk the angular momentum due to the central
massive object must be taken into account together with the angular
momentum due to the cyclotron motion.

Considering the differential rotation under the Hill coordinate
[\cite{Hill1878}]  we obtain a second order ordinary differential equation
which degenerates to the one introduced in the equation (16) of
\cite{Harris1962} in the limit of $\Omega_{0} \rightarrow 0$. The solution
cannot be expressed with a superposition of simple functions. Therefore
we have simply adopted a superposition of a conventional Harris solution
and a Keplerian differential rotation as an initial asymptotic
solution. Here we investigate whether this approximation is valid
or not in the case of $\Omega_{0} / \ocyci \ll 1$. We carry out
five comparative runs with no initial trigger for the magnetic
reconnection, i.e. $B_{1}=0$ in Equation (\ref{eq_PertB}) and $\eta_{\rm c}=0$
in Equation (\ref{eq_resistivity}), with the rotational parameter used in the
main runs. We also restrict the vertical size of the simulation
domain ($L_{z}$) to guarantee the outside of the current sheet region to be
stable to the MRI. From the results, it is confirmed that the initial
structure of the current sheet is conserved in time. However, we also
find that the neutral point of the current sheet moves in time. 

\begin{figure}[htbp]
 \begin{center}
 \includegraphics[bb=20 20 1000 760,
  width=\hsize]{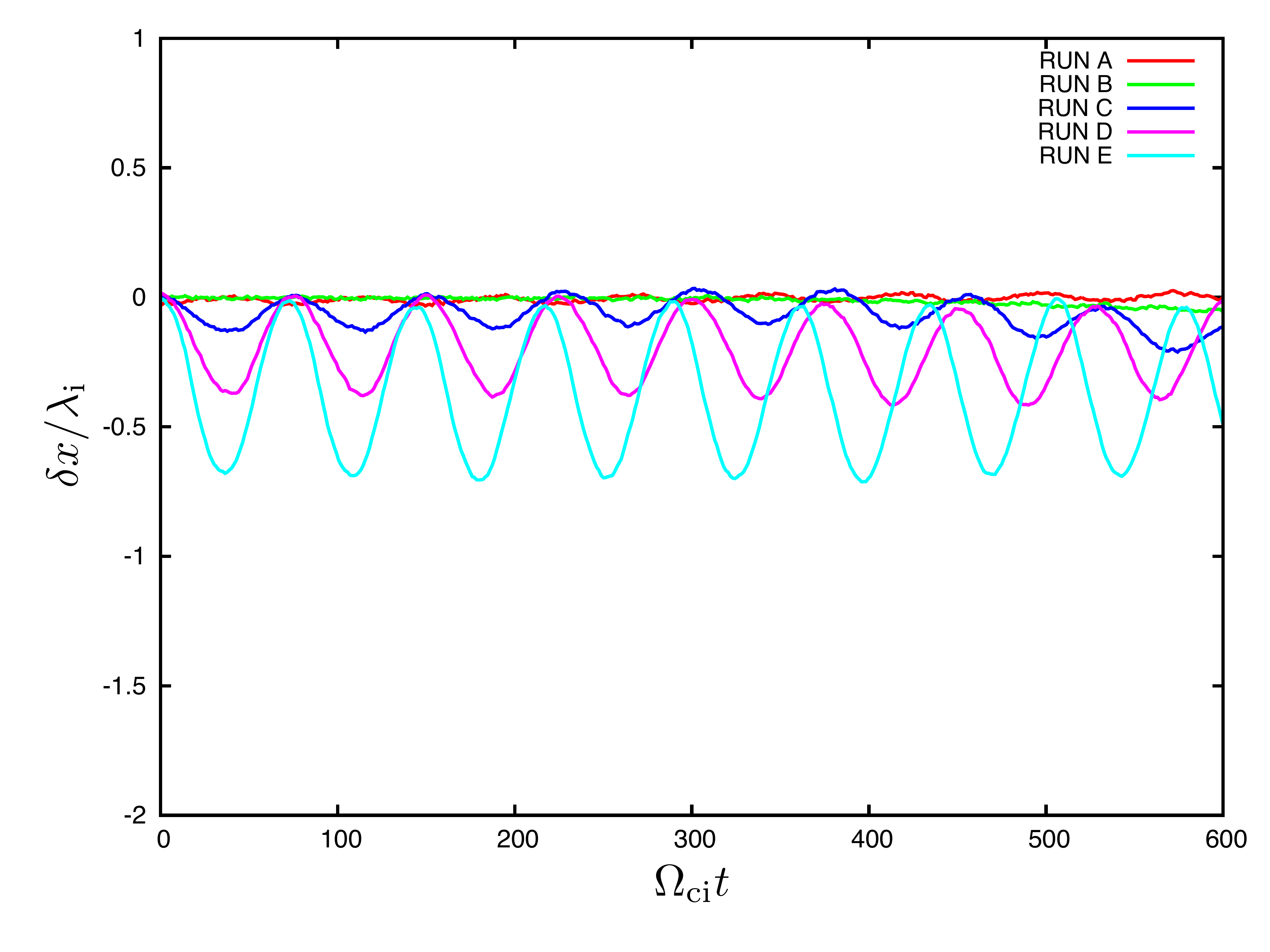}
 \end{center}
 \caption{The oscillation of the neutral point. The deviation from the
 initial
 neutral point ($\delta x$) in the unit of inertia length is plotted in
 the vertical direction.}\label{fig_EqOsci}
\end{figure}
Figure \ref{fig_EqOsci} shows the position of the neutral
point observed in each calculation. As the rotational parameter
increases the amplitude of
oscillation becomes significant though its amplitude is smaller than the
initial width of the current sheet. This is due
to the initial current in the neutral sheet. Since in this calculation, the
ions are hotter than the electrons, the dominant component with
subject to an out-of-plane initial current is the ions. As the ions have
finite mass in the hybrid framework, any initial motion inside the
equatorial plane will couple to the differential rotation through the 
Coriolis force. In this case the 
initial out of plane drift $\pm V_{y}$ creates $\mp V_{x}$ with $\pi/2$
phase delays due to the Coriolis force. Once $V_{x}$ has created the
compression and the expansion of the plasma and the magnetic field
take
place since the direction of the drift motion is opposite with the each
current sheet. Therefore, the oscillation can roughly be regarded as a
perpendicular mode of the fast mode magnetosonic wave. The estimated
frequency of perpendicular fast mode wave in this system is
approximately $0.1 \ocyci$, which roughly explains the results.

As the amplitude of this oscillation is relatively small
throughout all the parameters used in this calculation we find it is
valid to use a Harris solution as an approximate equilibrium solution in
this
system provided that the rotational parameter $\Omega_{0} / \ocyci$ is
sufficiently small.
However, since the anomalous resistivity is imposed locally in the
current sheet center, and since the current sheet oscillates in time,
one must pay attention to the detection of the current sheet center.
At each time step the current sheet centers ($x_{\rm
c}(t)$ and  $x'_{\rm c} (t)$) are detected by calculating
a local maximum of $\vlne \left(\vec{\nabla \times
B}\right)_{y} \vlne$ in the vicinity of the current sheet centers
which has also been detected in the previous time step.

It is worth noting that the initial direction of $V_{x}$ generated by
the neutral sheet current and the direction of the X-point migration
introduced in Sec. \ref{sec_results} is opposite. In addition, we have
also confirmed, by setting $\eta_{\rm c}=0$ in equation
(\ref{eq_resistivity}), that the way of detecting the current sheet
center does not affect the direction of the X-point migration.
Therefore, we conclude that the X-point migration is independent to
an initial plasma sheet oscillation and purely a result of coupling with
the structure of the reconnection and the differential rotation.


\providecommand{\noopsort}[1]{}\providecommand{\singleletter}[1]{#1}
%


\begin{thebibliography}{33}%
\makeatletter
\providecommand \@ifxundefined [1]{%
 \@ifx{#1\undefined}
}%
\providecommand \@ifnum [1]{%
 \ifnum #1\expandafter \@firstoftwo
 \else \expandafter \@secondoftwo
 \fi
}%
\providecommand \@ifx [1]{%
 \ifx #1\expandafter \@firstoftwo
 \else \expandafter \@secondoftwo
 \fi
}%
\providecommand \natexlab [1]{#1}%
\providecommand \enquote  [1]{``#1''}%
\providecommand \bibnamefont  [1]{#1}%
\providecommand \bibfnamefont [1]{#1}%
\providecommand \citenamefont [1]{#1}%
\providecommand \href@noop [0]{\@secondoftwo}%
\providecommand \href [0]{\begingroup \@sanitize@url \@href}%
\providecommand \@href[1]{\@@startlink{#1}\@@href}%
\providecommand \@@href[1]{\endgroup#1\@@endlink}%
\providecommand \@sanitize@url [0]{\catcode `\\12\catcode `\$12\catcode
  `\&12\catcode `\#12\catcode `\^12\catcode `\_12\catcode `\%12\relax}%
\providecommand \@@startlink[1]{}%
\providecommand \@@endlink[0]{}%
\providecommand \url  [0]{\begingroup\@sanitize@url \@url }%
\providecommand \@url [1]{\endgroup\@href {#1}{\urlprefix }}%
\providecommand \urlprefix  [0]{URL }%
\providecommand \Eprint [0]{\href }%
\providecommand \doibase [0]{http://dx.doi.org/}%
\providecommand \selectlanguage [0]{\@gobble}%
\providecommand \bibinfo  [0]{\@secondoftwo}%
\providecommand \bibfield  [0]{\@secondoftwo}%
\providecommand \translation [1]{[#1]}%
\providecommand \BibitemOpen [0]{}%
\providecommand \bibitemStop [0]{}%
\providecommand \bibitemNoStop [0]{.\EOS\space}%
\providecommand \EOS [0]{\spacefactor3000\relax}%
\providecommand \BibitemShut  [1]{\csname bibitem#1\endcsname}%
\let\auto@bib@innerbib\@empty
\bibitem [{\citenamefont {{Arzner}}\ and\ \citenamefont {{Scholer}}(2001)}]{ArznerScholerJGR2001}%
  \BibitemOpen
  \bibfield  {author} {\bibinfo {author} {\bibnamefont {{Arzner}},
  \bibfnamefont {K.}}\ and\ \bibinfo {author} {\bibnamefont {{Scholer}},
  \bibfnamefont {M.}},\ }\bibfield  {title} {\enquote {\bibinfo {title}
  {{Kinetic structure of the post plasmoid plasma sheet during magnetotail
  reconnection}},}\ }\href {\doibase 10.1029/2000JA000179} {\bibfield
  {journal} {\bibinfo  {journal} {Journal of Geophysical Research}\ }\textbf
  {\bibinfo {volume} {106}},\ \bibinfo {pages} {3827--3844} (\bibinfo {year}
  {2001})}%
\bibitem [{\citenamefont {{Asano}}\ \emph {et~al.}(2003)\citenamefont
  {{Asano}}, \citenamefont {{Mukai}}, \citenamefont {{Hoshino}}, \citenamefont
  {{Saito}}, \citenamefont {{Hayakawa}},\ and\ \citenamefont
  {{Nagai}}}]{Asano+2003}%
  \BibitemOpen
  \bibfield  {author} {\bibinfo {author} {\bibnamefont {{Asano}}, \bibfnamefont
  {Y.}}, \bibinfo {author} {\bibnamefont {{Mukai}}, \bibfnamefont {T.}},
  \bibinfo {author} {\bibnamefont {{Hoshino}}, \bibfnamefont {M.}}, \bibinfo
  {author} {\bibnamefont {{Saito}}, \bibfnamefont {Y.}}, \bibinfo {author}
  {\bibnamefont {{Hayakawa}}, \bibfnamefont {H.}}, \ and\ \bibinfo {author}
  {\bibnamefont {{Nagai}}, \bibfnamefont {T.}},\ }\bibfield  {title} {\enquote
  {\bibinfo {title} {{Evolution of the thin current sheet in a substorm
  observed by Geotail}},}\ }\href {\doibase 10.1029/2002JA009785} {\bibfield
  {journal} {\bibinfo  {journal} {Journal of Geophysical Research}\ }\textbf
  {\bibinfo {volume} {108}},\ \bibinfo {eid} {1189} (\bibinfo {year}
  {2003})}%
\bibitem [{\citenamefont {{Balbus}}\ and\ \citenamefont
  {{Hawley}}(1991)}]{BalbusHawley1991A}%
  \BibitemOpen
  \bibfield  {author} {\bibinfo {author} {\bibnamefont {{Balbus}},
  \bibfnamefont {S.~A.}}\ and\ \bibinfo {author} {\bibnamefont {{Hawley}},
  \bibfnamefont {J.~F.}},\ }\bibfield  {title} {\enquote {\bibinfo {title} {{A
  powerful local shear instability in weakly magnetized disks. I - Linear
  analysis. II - Nonlinear evolution}},}\ }\href {\doibase 10.1086/170270}
  {\bibfield  {journal} {\bibinfo  {journal} {$\apj$}\ }\textbf {\bibinfo
  {volume} {376}},\ \bibinfo {pages} {214--233} (\bibinfo {year}
  {1991})}%
\bibitem [{\citenamefont {{Balbus}}\ and\ \citenamefont
  {{Hawley}}(1998)}]{BH1998Review}%
  \BibitemOpen
  \bibfield  {author} {\bibinfo {author} {\bibnamefont {{Balbus}},
  \bibfnamefont {S.~A.}}\ and\ \bibinfo {author} {\bibnamefont {{Hawley}},
  \bibfnamefont {J.~F.}},\ }\bibfield  {title} {\enquote {\bibinfo {title}
  {{Instability, turbulence, and enhanced transport in accretion disks}},}\
  }\href {\doibase 10.1103/RevModPhys.70.1} {\bibfield  {journal} {\bibinfo
  {journal} {Reviews of Modern Physics}\ }\textbf {\bibinfo {volume} {70}},\
  \bibinfo {pages} {1--53} (\bibinfo {year} {1998})}%
\bibitem [{\citenamefont {{Balbus}}\ and\ \citenamefont
  {{Terquem}}(2001)}]{BalbusTerquem2001}%
  \BibitemOpen
  \bibfield  {author} {\bibinfo {author} {\bibnamefont {{Balbus}},
  \bibfnamefont {S.~A.}}\ and\ \bibinfo {author} {\bibnamefont {{Terquem}},
  \bibfnamefont {C.}},\ }\bibfield  {title} {\enquote {\bibinfo {title}
  {{Linear Analysis of the Hall Effect in Protostellar Disks}},}\ }\href
  {\doibase 10.1086/320452} {\bibfield  {journal} {\bibinfo  {journal} {$\apj$}\
  }\textbf {\bibinfo {volume} {552}},\ \bibinfo {pages} {235--247} (\bibinfo
  {year} {2001})}%
\bibitem [{\citenamefont {{Chandrasekhar}}(1960)}]{Chandrasekhar1960}%
  \BibitemOpen
  \bibfield  {author} {\bibinfo {author} {\bibnamefont {{Chandrasekhar}},
  \bibfnamefont {S.}},\ }\bibfield  {title} {\enquote {\bibinfo {title} {{The
  Stability of Non-Dissipative Couette Flow in Hydromagnetics}},}\ }\href
  {\doibase 10.1073/pnas.46.2.253} {\bibfield  {journal} {\bibinfo  {journal}
  {Proceedings of the National Academy of Science}\ }\textbf {\bibinfo {volume}
  {46}},\ \bibinfo {pages} {253--257} (\bibinfo {year} {1960})}%
\bibitem [{\citenamefont {{Chew}}, \citenamefont {{Goldberger}},\ and\
  \citenamefont {{Low}}(1956)}]{CGL1956}%
  \BibitemOpen
  \bibfield  {author} {\bibinfo {author} {\bibnamefont {{Chew}}, \bibfnamefont
  {G.~F.}}, \bibinfo {author} {\bibnamefont {{Goldberger}}, \bibfnamefont
  {M.~L.}}, \ and\ \bibinfo {author} {\bibnamefont {{Low}}, \bibfnamefont
  {F.~E.}},\ }\bibfield  {title} {\enquote {\bibinfo {title} {{The Boltzmann
  Equation and the One-Fluid Hydromagnetic Equations in the Absence of Particle
  Collisions}},}\ }\href {\doibase 10.1098/rspa.1956.0116} {\bibfield
  {journal} {\bibinfo  {journal} {Royal Society of London Proceedings Series
  A}\ }\textbf {\bibinfo {volume} {236}},\ \bibinfo {pages} {112--118}
  (\bibinfo {year} {1956})}%
\bibitem [{\citenamefont {{Ferraro}}(2007)}]{Ferraro2007}%
  \BibitemOpen
  \bibfield  {author} {\bibinfo {author} {\bibnamefont {{Ferraro}},
  \bibfnamefont {N.~M.}},\ }\bibfield  {title} {\enquote {\bibinfo {title}
  {{Finite Larmor Radius Effects on the Magnetorotational Instability}},}\
  }\href {\doibase 10.1086/517877} {\bibfield  {journal} {\bibinfo  {journal}
  {$\apj$}\ }\textbf {\bibinfo {volume} {662}},\ \bibinfo {pages} {512--516}
  (\bibinfo {year} {2007})}%
\bibitem [{\citenamefont {{Goodman}}\ and\ \citenamefont
  {{Xu}}(1994)}]{GoodmanXu1994}%
  \BibitemOpen
  \bibfield  {author} {\bibinfo {author} {\bibnamefont {{Goodman}},
  \bibfnamefont {J.}}\ and\ \bibinfo {author} {\bibnamefont {{Xu}},
  \bibfnamefont {G.}},\ }\bibfield  {title} {\enquote {\bibinfo {title}
  {{Parasitic instabilities in magnetized, differentially rotating disks}},}\
  }\href {\doibase 10.1086/174562} {\bibfield  {journal} {\bibinfo  {journal}
  {$\apj$}\ }\textbf {\bibinfo {volume} {432}},\ \bibinfo {pages} {213--223}
  (\bibinfo {year} {1994})}%
\bibitem [{\citenamefont {Harris}(1962)}]{Harris1962}%
  \BibitemOpen
  \bibfield  {author} {\bibinfo {author} {\bibnamefont {Harris}, \bibfnamefont
  {E.~G.}},\ }\bibfield  {title} {\enquote {\bibinfo {title} {{On a plasma
  sheath separating regions of oppositely directed magnetic field}},}\ }\href
  {\doibase 10.1007/BF02733547} {\bibfield  {journal} {\bibinfo  {journal} {Il
  Nuovo Cimento}\ }\textbf {\bibinfo {volume} {23}},\ \bibinfo {pages}
  {115--121} (\bibinfo {year} {1962})}%
\bibitem [{\citenamefont {{Hau}}\ and\ \citenamefont
  {{Sonnerup}}(1992)}]{HauSonnerup1992}%
  \BibitemOpen
  \bibfield  {author} {\bibinfo {author} {\bibnamefont {{Hau}}, \bibfnamefont
  {L.-N.}}\ and\ \bibinfo {author} {\bibnamefont {{Sonnerup}}, \bibfnamefont
  {B.~U.~O.}},\ }\bibfield  {title} {\enquote {\bibinfo {title} {{The thickness
  of resistive-dispersive shocks}},}\ }\href {\doibase 10.1029/92JA00138}
  {\bibfield  {journal} {\bibinfo  {journal} {Journal of Geophysical Research}\
  }\textbf {\bibinfo {volume} {97}},\ \bibinfo {pages} {8269--8275} (\bibinfo
  {year} {1992})}%
\bibitem [{\citenamefont {{Hawley}}\ and\ \citenamefont
  {{Balbus}}(1991)}]{BalbusHawley1991B}%
  \BibitemOpen
  \bibfield  {author} {\bibinfo {author} {\bibnamefont {{Hawley}},
  \bibfnamefont {J.~F.}}\ and\ \bibinfo {author} {\bibnamefont {{Balbus}},
  \bibfnamefont {S.~A.}},\ }\bibfield  {title} {\enquote {\bibinfo {title} {{A
  Powerful Local Shear Instability in Weakly Magnetized Disks. II. Nonlinear
  Evolution}},}\ }\href {\doibase 10.1086/170271} {\bibfield  {journal}
  {\bibinfo  {journal} {$\apj$}\ }\textbf {\bibinfo {volume} {376}},\ \bibinfo
  {pages} {223} (\bibinfo {year} {1991})}%
\bibitem [{\citenamefont {{Hawley}}\ and\ \citenamefont
  {{Balbus}}(1992)}]{BalbusHawley1991C}%
  \BibitemOpen
  \bibfield  {author} {\bibinfo {author} {\bibnamefont {{Hawley}},
  \bibfnamefont {J.~F.}}\ and\ \bibinfo {author} {\bibnamefont {{Balbus}},
  \bibfnamefont {S.~A.}},\ }\bibfield  {title} {\enquote {\bibinfo {title} {{A
  powerful local shear instability in weakly magnetized disks. III - Long-term
  evolution in a shearing sheet}},}\ }\href {\doibase 10.1086/172021}
  {\bibfield  {journal} {\bibinfo  {journal} {$\apj$}\ }\textbf {\bibinfo
  {volume} {400}},\ \bibinfo {pages} {595--609} (\bibinfo {year}
  {1992})}%
\bibitem [{\citenamefont {{Hawley}}, \citenamefont {{Gammie}},\ and\
  \citenamefont {{Balbus}}(1995)}]{Hawley+1995}%
  \BibitemOpen
  \bibfield  {author} {\bibinfo {author} {\bibnamefont {{Hawley}},
  \bibfnamefont {J.~F.}}, \bibinfo {author} {\bibnamefont {{Gammie}},
  \bibfnamefont {C.~F.}}, \ and\ \bibinfo {author} {\bibnamefont {{Balbus}},
  \bibfnamefont {S.~A.}},\ }\bibfield  {title} {\enquote {\bibinfo {title}
  {{Local Three-dimensional Magnetohydrodynamic Simulations of Accretion
  Disks}},}\ }\href {\doibase 10.1086/175311} {\bibfield  {journal} {\bibinfo
  {journal} {$\apj$}\ }\textbf {\bibinfo {volume} {440}},\ \bibinfo {pages} {742}
  (\bibinfo {year} {1995})}%
\bibitem [{\citenamefont {{Hesse}}\ and\ \citenamefont
  {{Winske}}(1993)}]{HesseWinske1993GRL}%
  \BibitemOpen
  \bibfield  {author} {\bibinfo {author} {\bibnamefont {{Hesse}}, \bibfnamefont
  {M.}}\ and\ \bibinfo {author} {\bibnamefont {{Winske}}, \bibfnamefont {D.}},\
  }\bibfield  {title} {\enquote {\bibinfo {title} {{Hybrid simulations of
  collisionless ion tearing}},}\ }\href {\doibase 10.1029/93GL01250} {\bibfield
   {journal} {\bibinfo  {journal} {Geophysical Research Letters}\ }\textbf
  {\bibinfo {volume} {20}},\ \bibinfo {pages} {1207--1210} (\bibinfo {year}
  {1993})}%
\bibitem [{\citenamefont {{Higashimori}}\ and\ \citenamefont
  {{Hoshino}}(2012)}]{HigashimoriHoshino2012}%
  \BibitemOpen
  \bibfield  {author} {\bibinfo {author} {\bibnamefont {{Higashimori}},
  \bibfnamefont {K.}}\ and\ \bibinfo {author} {\bibnamefont {{Hoshino}},
  \bibfnamefont {M.}},\ }\bibfield  {title} {\enquote {\bibinfo {title} {{The
  relation between ion temperature anisotropy and formation of slow shocks in
  collisionless magnetic reconnection}},}\ }\href {\doibase
  10.1029/2011JA016817} {\bibfield  {journal} {\bibinfo  {journal} {Journal of
  Geophysical Research (Space Physics)}\ }\textbf {\bibinfo {volume} {117}},\
  \bibinfo {eid} {A01220} (\bibinfo {year} {2012})}%
\bibitem [{\citenamefont {Hill}(1878)}]{Hill1878}%
  \BibitemOpen
  \bibfield  {author} {\bibinfo {author} {\bibnamefont {Hill}, \bibfnamefont
  {G.~W.}},\ }\bibfield  {title} {\enquote {\bibinfo {title} {Researches in the
  lunar theory},}\ }\href {http://www.jstor.org/stable/2369430} {\bibfield
  {journal} {\bibinfo  {journal} {American Journal of Mathematics}\ }\textbf
  {\bibinfo {volume} {1}},\ \bibinfo {pages} {pp. 5--26} (\bibinfo {year}
  {1878})}%
\bibitem [{\citenamefont {Horowitz}, \citenamefont {Shumaker},\ and\
  \citenamefont {Anderson}(1989)}]{Horowitz1989}%
  \BibitemOpen
  \bibfield  {author} {\bibinfo {author} {\bibnamefont {Horowitz},
  \bibfnamefont {E.~J.}}, \bibinfo {author} {\bibnamefont {Shumaker},
  \bibfnamefont {D.~E.}}, \ and\ \bibinfo {author} {\bibnamefont {Anderson},
  \bibfnamefont {D.~V.}},\ }\bibfield  {title} {\enquote {\bibinfo {title}
  {Qn3d: A three-dimensional quasi-neutral hybrid particle-in-cell code with
  applications to the tilt mode instability in field reversed
  configurations},}\ }\href {\doibase
  http://dx.doi.org/10.1016/0021-9991(89)90234-9} {\bibfield  {journal}
  {\bibinfo  {journal} {Journal of Computational Physics}\ }\textbf {\bibinfo
  {volume} {84}},\ \bibinfo {pages} {279 -- 310} (\bibinfo {year}
  {1989})}%
\bibitem [{\citenamefont {Hoshino}(2013)}]{Hoshino2013}%
  \BibitemOpen
  \bibfield  {author} {\bibinfo {author} {\bibnamefont {Hoshino}, \bibfnamefont
  {M.}},\ }\bibfield  {title} {\enquote {\bibinfo {title} {Particle
  acceleration during magnetorotational instability in a collisionless
  accretion disk},}\ }\href {http://stacks.iop.org/0004-637X/773/i=2/a=118}
  {\bibfield  {journal} {\bibinfo  {journal} {$\apj$}\ }\textbf {\bibinfo
  {volume} {773}},\ \bibinfo {pages} {118} (\bibinfo {year}
  {2013})}%
\bibitem [{\citenamefont {{Kulsrud}}(1983)}]{Kulsrud1983Text}%
  \BibitemOpen
  \bibfield  {author} {\bibinfo {author} {\bibnamefont {{Kulsrud}},
  \bibfnamefont {R.~M.}},\ }\bibfield  {title} {\enquote {\bibinfo {title}
  {{MHD description of plasma}},}\ }in\ \href@noop {} {\emph {\bibinfo
  {booktitle} {Basic Plasma Physics: Selected Chapters, Handbook of Plasma
  Physics, Volume 1}}},\ \bibinfo {editor} {edited by\ \bibinfo {editor}
  {\bibfnamefont {A.~A.}\ \bibnamefont {{Galeev}}}\ and\ \bibinfo {editor}
  {\bibfnamefont {R.~N.}\ \bibnamefont {{Sudan}}}}\ (\bibinfo {year}
	  {1983})\ (\bibinfo {publisher} {Amsterdam$\cl$North-Holland})
  p.~\bibinfo {pages} {1}%
\bibitem [{\citenamefont {{Matsumoto}}\ and\ \citenamefont
  {{Tajima}}(1995)}]{MatsumotoTajima1995}%
  \BibitemOpen
  \bibfield  {author} {\bibinfo {author} {\bibnamefont {{Matsumoto}},
  \bibfnamefont {R.}}\ and\ \bibinfo {author} {\bibnamefont {{Tajima}},
  \bibfnamefont {T.}},\ }\bibfield  {title} {\enquote {\bibinfo {title}
  {{Magnetic viscosity by localized shear flow instability in magnetized
  accretion disks}},}\ }\href {\doibase 10.1086/175739} {\bibfield  {journal}
  {\bibinfo  {journal} {$\apj$}\ }\textbf {\bibinfo {volume} {445}},\ \bibinfo
  {pages} {767--779} (\bibinfo {year} {1995})}%
\bibitem [{\citenamefont {{Narayan}}, \citenamefont {{Yi}},\ and\ \citenamefont
  {{Mahadevan}}(1995)}]{NarayanYiNature1995}%
  \BibitemOpen
  \bibfield  {author} {\bibinfo {author} {\bibnamefont {{Narayan}},
  \bibfnamefont {R.}}, \bibinfo {author} {\bibnamefont {{Yi}}, \bibfnamefont
  {I.}}, \ and\ \bibinfo {author} {\bibnamefont {{Mahadevan}}, \bibfnamefont
  {R.}},\ }\bibfield  {title} {\enquote {\bibinfo {title} {{Explaining the
  spectrum of Sagittarius A$^{*}$ with a model of an accreting black hole}},}\
  }\href {\doibase 10.1038/374623a0} {\bibfield  {journal} {\bibinfo  {journal}
  {$\nat$}\ }\textbf {\bibinfo {volume} {374}},\ \bibinfo {pages} {623--625}
  (\bibinfo {year} {1995})}%
\bibitem [{\citenamefont {{Quataert}}, \citenamefont {{Dorland}},\ and\
  \citenamefont {{Hammett}}(2002)}]{Quataert+2002}%
  \BibitemOpen
  \bibfield  {author} {\bibinfo {author} {\bibnamefont {{Quataert}},
  \bibfnamefont {E.}}, \bibinfo {author} {\bibnamefont {{Dorland}},
  \bibfnamefont {W.}}, \ and\ \bibinfo {author} {\bibnamefont {{Hammett}},
  \bibfnamefont {G.~W.}},\ }\bibfield  {title} {\enquote {\bibinfo {title}
  {{The Magnetorotational Instability in a Collisionless Plasma}},}\ }\href
  {\doibase 10.1086/342174} {\bibfield  {journal} {\bibinfo  {journal} {$\apj$}\
  }\textbf {\bibinfo {volume} {577}},\ \bibinfo {pages} {524--533} (\bibinfo
  {year} {2002})}%
\bibitem [{\citenamefont {{Riquelme}}\ \emph {et~al.}(2012)\citenamefont
  {{Riquelme}}, \citenamefont {{Quataert}}, \citenamefont {{Sharma}},\ and\
  \citenamefont {{Spitkovsky}}}]{Riquelme+2012}%
  \BibitemOpen
  \bibfield  {author} {\bibinfo {author} {\bibnamefont {{Riquelme}},
  \bibfnamefont {M.~A.}}, \bibinfo {author} {\bibnamefont {{Quataert}},
  \bibfnamefont {E.}}, \bibinfo {author} {\bibnamefont {{Sharma}},
  \bibfnamefont {P.}}, \ and\ \bibinfo {author} {\bibnamefont {{Spitkovsky}},
  \bibfnamefont {A.}},\ }\bibfield  {title} {\enquote {\bibinfo {title} {{Local
  Two-dimensional Particle-in-cell Simulations of the Collisionless
  Magnetorotational Instability}},}\ }\href {\doibase
  10.1088/0004-637X/755/1/50} {\bibfield  {journal} {\bibinfo  {journal}
  {$\apj$}\ }\textbf {\bibinfo {volume} {755}},\ \bibinfo {eid} {50} (\bibinfo
  {year} {2012})}%
\bibitem [{\citenamefont {{Sano}}\ and\ \citenamefont
  {{Inutsuka}}(2001)}]{SanoInutsuka2001}%
  \BibitemOpen
  \bibfield  {author} {\bibinfo {author} {\bibnamefont {{Sano}}, \bibfnamefont
  {T.}}\ and\ \bibinfo {author} {\bibnamefont {{Inutsuka}}, \bibfnamefont
  {S.-I.}},\ }\bibfield  {title} {\enquote {\bibinfo {title} {{Saturation and
  Thermalization of the Magnetorotational Instability: Recurrent Channel Flows
  and Reconnections}},}\ }\href {\doibase 10.1086/324763} {\bibfield  {journal}
  {\bibinfo  {journal} {Astrophys. J. Let.}\ }\textbf {\bibinfo {volume}
  {561}},\ \bibinfo {pages} {L179--L182} (\bibinfo {year} {2001})}%
\bibitem [{\citenamefont {{Sano}}\ and\ \citenamefont
  {{Stone}}(2002)}]{SanoStone2002A}%
  \BibitemOpen
  \bibfield  {author} {\bibinfo {author} {\bibnamefont {{Sano}}, \bibfnamefont
  {T.}}\ and\ \bibinfo {author} {\bibnamefont {{Stone}}, \bibfnamefont
  {J.~M.}},\ }\bibfield  {title} {\enquote {\bibinfo {title} {{The Effect of
  the Hall Term on the Nonlinear Evolution of the Magnetorotational
  Instability. I. Local Axisymmetric Simulations}},}\ }\href {\doibase
  10.1086/339504} {\bibfield  {journal} {\bibinfo  {journal} {$\apj$}\ }\textbf
  {\bibinfo {volume} {570}},\ \bibinfo {pages} {314--328} (\bibinfo {year}
  {2002})}%
\bibitem [{\citenamefont {{Shakura}}\ and\ \citenamefont
  {{Sunyaev}}(1973)}]{ShakuraSunyaev1973}%
  \BibitemOpen
  \bibfield  {author} {\bibinfo {author} {\bibnamefont {{Shakura}},
  \bibfnamefont {N.~I.}}\ and\ \bibinfo {author} {\bibnamefont {{Sunyaev}},
  \bibfnamefont {R.~A.}},\ }\bibfield  {title} {\enquote {\bibinfo {title}
  {{Black holes in binary systems. Observational appearance.}}}\ }\href@noop {}
  {\bibfield  {journal} {\bibinfo  {journal} {Astron. \& Astrophys.}\ }\textbf
  {\bibinfo {volume} {24}},\ \bibinfo {pages} {337--355} (\bibinfo {year}
  {1973})}%
\bibitem [{\citenamefont {{Sharma}}\ \emph {et~al.}(2006)\citenamefont
  {{Sharma}}, \citenamefont {{Hammett}}, \citenamefont {{Quataert}},\ and\
  \citenamefont {{Stone}}}]{Sharma+2006}%
  \BibitemOpen
  \bibfield  {author} {\bibinfo {author} {\bibnamefont {{Sharma}},
  \bibfnamefont {P.}}, \bibinfo {author} {\bibnamefont {{Hammett}},
  \bibfnamefont {G.~W.}}, \bibinfo {author} {\bibnamefont {{Quataert}},
  \bibfnamefont {E.}}, \ and\ \bibinfo {author} {\bibnamefont {{Stone}},
  \bibfnamefont {J.~M.}},\ }\bibfield  {title} {\enquote {\bibinfo {title}
  {{Shearing Box Simulations of the MRI in a Collisionless Plasma}},}\ }\href
  {\doibase 10.1086/498405} {\bibfield  {journal} {\bibinfo  {journal} {$\apj$}\
  }\textbf {\bibinfo {volume} {637}},\ \bibinfo {pages} {952--967} (\bibinfo
  {year} {2006})}%
\bibitem [{\citenamefont {{Snyder}}, \citenamefont {{Hammett}},\ and\
  \citenamefont {{Dorland}}(1997)}]{Snyder1997}%
  \BibitemOpen
  \bibfield  {author} {\bibinfo {author} {\bibnamefont {{Snyder}},
  \bibfnamefont {P.~B.}}, \bibinfo {author} {\bibnamefont {{Hammett}},
  \bibfnamefont {G.~W.}}, \ and\ \bibinfo {author} {\bibnamefont {{Dorland}},
  \bibfnamefont {W.}},\ }\bibfield  {title} {\enquote {\bibinfo {title}
  {{Landau fluid models of collisionless magnetohydrodynamics}},}\ }\href
  {\doibase 10.1063/1.872517} {\bibfield  {journal} {\bibinfo  {journal}
  {Physics of Plasmas}\ }\textbf {\bibinfo {volume} {4}},\ \bibinfo {pages}
  {3974--3985} (\bibinfo {year} {1997})}%
\bibitem [{\citenamefont {{Sonnerup}}(1979)}]{Sonnerup1979Text}%
  \BibitemOpen
  \bibfield  {author} {\bibinfo {author} {\bibnamefont {{Sonnerup}},
  \bibfnamefont {B.~U.~{\"O}.}},\ }\enquote {\bibinfo {title} {{Magnetic field
  reconnection}},}\ in\ \href@noop {} {\emph {\bibinfo {booktitle} {Solar
  System Plasma Physics}}},\ \bibinfo {editor} {edited by\ \bibinfo {editor}
  {\bibfnamefont {L.~J.}\ \bibnamefont {{Lanzerotti}}}, \bibinfo {editor}
  {\bibfnamefont {C.~F.}\ \bibnamefont {{Kennel}}}, \ and\ \bibinfo {editor}
  {\bibfnamefont {E.~N.}\ \bibnamefont {{Parker}}}}\ (\bibinfo {year}
	  {1979})\ (\bibinfo {publisher} {Amsterdam$\cl$North-Holland})
  pp.\ \bibinfo {pages} {45--108}%
\bibitem [{\citenamefont {{Terasawa}}(1983)}]{Terasawa1983}%
  \BibitemOpen
  \bibfield  {author} {\bibinfo {author} {\bibnamefont {{Terasawa}},
  \bibfnamefont {T.}},\ }\bibfield  {title} {\enquote {\bibinfo {title} {{Hall
  current effect on tearing mode instability}},}\ }\href {\doibase
  10.1029/GL010i006p00475} {\bibfield  {journal} {\bibinfo  {journal}
  {Geophysical Research Letters}\ }\textbf {\bibinfo {volume} {10}},\ \bibinfo
  {pages} {475--478} (\bibinfo {year} {1983})}%
\bibitem [{\citenamefont {Velikhov}(1959)}]{Velikhov1959}%
  \BibitemOpen
  \bibfield  {author} {\bibinfo {author} {\bibnamefont {Velikhov},
  \bibfnamefont {E.}},\ }\bibfield  {title} {\enquote {\bibinfo {title}
  {Stability of an ideally conducting liquid flowing between rotating cylinders
  in a magnetic field},}\ }\href
  {http://www.osti.gov/scitech/servlets/purl/4232891} {\bibfield  {journal}
  {\bibinfo  {journal} {Journal Name: Zhur. Eksptlʼ. i Teoret. Fiz.}\
	  }\textbf {\bibinfo
  {volume} {36}},\ \bibinfo {pages} {1398}
  (\bibinfo {year} {1959})}%
\bibitem [{\citenamefont {{Zenitani}}\ \emph {et~al.}(2011)\citenamefont
  {{Zenitani}}, \citenamefont {{Hesse}}, \citenamefont {{Klimas}},
  \citenamefont {{Black}},\ and\ \citenamefont {{Kuznetsova}}}]{Zenitani+2011}%
  \BibitemOpen
  \bibfield  {author} {\bibinfo {author} {\bibnamefont {{Zenitani}},
  \bibfnamefont {S.}}, \bibinfo {author} {\bibnamefont {{Hesse}}, \bibfnamefont
  {M.}}, \bibinfo {author} {\bibnamefont {{Klimas}}, \bibfnamefont {A.}},
  \bibinfo {author} {\bibnamefont {{Black}}, \bibfnamefont {C.}}, \ and\
  \bibinfo {author} {\bibnamefont {{Kuznetsova}}, \bibfnamefont {M.}},\
  }\bibfield  {title} {\enquote {\bibinfo {title} {{The inner structure of
  collisionless magnetic reconnection: The electron-frame dissipation measure
  and Hall fields}},}\ }\href {\doibase 10.1063/1.3662430} {\bibfield
  {journal} {\bibinfo  {journal} {Physics of Plasmas}\ }\textbf {\bibinfo
  {volume} {18}},\ \bibinfo {pages} {122108} (\bibinfo {year}
  {2011})}%
\end{thebibliography}

\end{document}